\documentclass[english,aps,pre,letter,superscriptaddress,preprint,linenumbers,nofootnoteinbib]{revtex4}
\usepackage[T1]{fontenc}
\usepackage[latin9]{inputenc}
\usepackage{xcolor}
\usepackage{pdfcolmk}
\usepackage{babel}
\usepackage{float}
\usepackage{amsmath}
\usepackage{amssymb}
\usepackage{graphicx}
\PassOptionsToPackage{normalem}{ulem}
\usepackage{ulem}
\usepackage[unicode=true,pdfusetitle,
 bookmarks=true,bookmarksnumbered=false,bookmarksopen=false,
 breaklinks=false,pdfborder={0 0 1},backref=false,colorlinks=false]
 {hyperref}

\makeatletter

\floatstyle{ruled}
\newfloat{algorithm}{tbp}{loa}
\providecommand{\algorithmname}{Algorithm}
\floatname{algorithm}{\protect\algorithmname}
\providecolor{lyxadded}{rgb}{1,0,0}
\providecolor{lyxdeleted}{rgb}{0,0,1}

\DeclareRobustCommand{\lyxsout}[1]{\ifx\\#1\else\sout{#1}\fi}

\@ifundefined{textcolor}{}
{%
 \definecolor{BLACK}{gray}{0}
 \definecolor{WHITE}{gray}{1}
 \definecolor{RED}{rgb}{1,0,0}
 \definecolor{GREEN}{rgb}{0,1,0}
 \definecolor{BLUE}{rgb}{0,0,1}
 \definecolor{CYAN}{cmyk}{1,0,0,0}
 \definecolor{MAGENTA}{cmyk}{0,1,0,0}
 \definecolor{YELLOW}{cmyk}{0,0,1,0}
}

\usepackage{ae,aecompl}

\makeatother

\begin{document}

\title{Efficient Reactive Brownian Dynamics}

\author{Aleksandar Donev}
\email{donev@courant.nyu.edu}

\selectlanguage{english}%

\affiliation{Courant Institute of Mathematical Sciences, New York University,
New York, NY 10012}

\author{Chiao-Yu Yang}

\affiliation{Department of Mathematics, UC Berkeley, Berkeley CA 94720}

\affiliation{Courant Institute of Mathematical Sciences, New York University,
New York, NY 10012}

\author{Changho Kim}

\affiliation{Computational Research Division, Lawrence Berkeley National Laboratory,
Berkeley CA, 94720}
\begin{abstract}
\textcolor{black}{We develop a }Split Reactive Brownian Dynamics (SRBD)
algorithm for particle simulations of reaction-diffusion systems based
on the Doi or volume reactivity model, in which pairs of particles
react with a specified Poisson rate if they are closer than a chosen
reactive distance. In our Doi model, we ensure that the microscopic
reaction rules for various association and dissociation reactions
are consistent with detailed balance (time reversibility) at thermodynamic
equilibrium. The SRBD algorithm uses Strang splitting in time to separate
reaction and diffusion, and solves both the diffusion-only and reaction-only
subproblems exactly, even at high packing densities. To efficiently
process reactions without uncontrolled approximations, SRBD employs
an event-driven algorithm that processes reactions in a time-ordered
sequence over the duration of the time step. A grid of cells with
size larger than all of the reactive distances is used to schedule
and process the reactions, but unlike traditional grid-based methods
such as Reaction-Diffusion Master Equation (RDME) algorithms, the
results of SRBD are statistically independent of the size of the grid
used to accelerate the processing of reactions. We use the SRBD algorithm
to compute the effective macroscopic reaction rate for both reaction-
and diffusion-limited irreversible association in three dimensions,
and compare to existing theoretical predictions at low and moderate
densities. We also study long-time tails in the time correlation functions
for reversible association at thermodynamic equilibrium, and compare
to recent theoretical predictions. Finally, we compare different particle
and continuum methods on a model exhibiting a Turing-like instability
and pattern formation. Our studies reinforce the common finding that
microscopic mechanisms and correlations matter for diffusion-limited
systems, making continuum and even mesoscopic modeling of such systems
difficult or impossible. We also find that for models in which particles
diffuse off lattice, such as the Doi model, reactions lead to a spurious
enhancement of the effective diffusion coefficients.
\end{abstract}
\maketitle
\global\long\def\V#1{\boldsymbol{#1}}
\global\long\def\M#1{\boldsymbol{#1}}
\global\long\def\Set#1{\mathbb{#1}}

\global\long\def\D#1{\Delta#1}
\global\long\def\d#1{\delta#1}

\global\long\def\norm#1{\left\Vert #1\right\Vert }
\global\long\def\abs#1{\left|#1\right|}

\global\long\def\grad{\M{\nabla}}
\global\long\def\avv#1{\langle#1\rangle}
\global\long\def\av#1{\left\langle #1\right\rangle }

\global\long\def\P{\mathcal{P}}

\global\long\def\ki{k}
\global\long\def\wi{\omega}

\global\long\def\bu{\V u}
 \global\long\def\bv{\V v}
 \global\long\def\br{\V r}

\global\long\def\sM#1{\M{\mathcal{#1}}}
\global\long\def\Mob{\sM M}
\global\long\def\J{\sM J}
\global\long\def\S{\sM S}
\global\long\def\L{\sM L}

\section{Introduction}

It is widely appreciated that fluctuations affect reactive systems
in important ways and should be retained, rather than averaged over,
in reaction-diffusion modeling. In stochastic biochemical systems,
such as reactions inside the cytoplasm, or in catalytic processes,
some of the reacting molecules are present in very small numbers and
therefore discrete stochastic models are necessary to describe the
system \cite{BiochemicalDiffusionReaction_Review,StochChemSpatialNature}.
In diffusion-limited reactive systems, such as simple coagulation
$2A\rightarrow A_{2}$ or annihilation $A+B\rightarrow\emptyset$,
spatial fluctuations in the concentration of the reactants grow as
the reaction progresses and must be accounted for to accurately model
the correct macroscopic behavior \cite{Annihilation_AplusB,Coagulation_Renormalization,LongTimeTails_ABC,AsymptoticPowerLaw_ABC}.
In unstable systems, such as diffusion-driven Turing instabilities
\cite{Turing_Fluctuations1,Turing_Fluctuations2,Turing_RDME,Turing_RDME_2,Turing_MD},
fluctuations are responsible for initiating the instability and have
been shown to profoundly affect the patterns in ways relevant to morphogenesis
\cite{Turing_Fluctuations1,Turing_Fluctuations2}. In systems with
a marginally-stable manifold, fluctuations lead to a drift along this
manifold that cannot be described by the traditional law of mass action;
this has been suggested as being an important mechanism in the emergence
of life \cite{MarginalStability_Brogioli,MarginalStability_Brogioli2,MarginalStability_LMA}.

Much of the work on modeling stochastic chemistry has been for homogeneous,
``well-mixed'' systems, but there is also a steady and growing interest
in spatial models \cite{BiochemicalDiffusionReaction_Review,StochChemSpatialNature,RDME_Review_Erban}.
Reaction-diffusion problems are often studied using the Reaction Diffusion
Master Equation (RDME) \cite{MalekRDME1975,KeizerBook,GardinerBook,vanKampen:07},
which extends the well-known Chemical Master Equation (CME) to spatially-varying
systems. In the RDME, the system is subdivided into reactive subvolumes
(cells) and diffusion is modeled as a discrete random walk by particles
hopping between cells, while reactions are modeled using CMEs local
to each cell \cite{RDME_Review_Erban}. A large number of efficient
and elaborate event-driven kinetic Monte Carlo algorithms for solving
the CME and RDME, exactly or approximately, have been developed with
many tracing their origins to the Stochastic Simulation Algorithm
(SSA) of Gillespie \cite{GillespieSSAreview}. Of particular importance
to the work presented here is the next subvolume method \cite{NextSubvolumeMethod_MesoRD},
which is an event-driven algorithm in which each cell independently
schedules the next reactive or diffusive event to take place inside
it. There are also a number of approximate techniques that speed the
Monte Carlo simulation of the RDME when there are many particles per
reactive cell, i.e., when the fluctuations are relatively weak. For
example, in multinomial diffusion algorithms \cite{MultinomialDiffusion_Gillespie,MultinomialDiffusion}
the diffusive hops between reactive cells are simulated by sampling
the number of jumps between neighboring cells using a multinomial
distribution. This can be combined with tau leaping for the reactions
using time splitting, as summarized in Appendix A in \cite{FluctReactDiff},
to give a method that does not simulate each individual event. One
can take this one step further toward the continuum limit by switching
to a real number rather than an integer representation for the number
of molecules in each cell, and use fluctuating hydrodynamics to simulate
diffusion \cite{FluctReactDiff}, giving a method that scales efficiently
to the deterministic reaction-diffusion limit.

The RDME, and related approaches based on local reactions inside each
reactive cell, have a number of important drawbacks, such as the lack
of convergence as the RDME grid is refined in the presence of binary
reactions \cite{MesoRD_GridResolution,RDME_Bimolecular_Petzold,CRDME}.
This means that one must choose the cells to be not too large, so
that spatial variations are resolved, nor too small, so that binary
reactions do not disappear. In fact, the common justification for
the RDME is to assume that each cell is well-mixed and homogeneous
\cite{GillespieSSAreview}, which is not true if one makes the cells
too small or too large. Variants of the RDME have been proposed that
improve or eliminate the lack of convergence as the grid is refined,
the most relevant to this work being the convergent RDME (CRDME) of
Isaacson \cite{CRDME}, in which reactions can happen between molecules
in neighboring cells as well. Another drawback of the RDME is the
fact that the CME, and therefore the RDME, require as input the \emph{macroscopic}
or \emph{mesoscopic} rates that enter in the law of mass action, rather
than \emph{microscopic} rates that define the reaction-diffusion process.
The conversion from the microscopic to the macroscopic rates is nontrivial
even for systems with only a single reaction, and even at very low
densities \cite{ReactionDiffusion_Doi,DoiModel_Erban}, let alone
at finite densities \cite{Coagulation_Renormalization_PRL,Coagulation_Renormalization}
or for systems with many species and reactions. It is \emph{not} correct,
in general, to use the macroscopic law of mass action coefficients
in the RDME, since this double counts fluctuations, as it is well-known
that the reaction rates are renormalized by spatial fluctuations \cite{Coagulation_Renormalization_PRL,Coagulation_Renormalization,RenormalizationReactionDiffusion_Review}.
The conversion of microscopic to \emph{mesoscopic} rates, i.e., rates
that depend on the cell size, is to our knowledge a completely open
problem. As clearly shown by the calculations of Erban and Chapman
\cite{DoiModel_Erban}, one can, alternatively, think of the RDME
as a \emph{microscopic }(rather than a mesoscopic or coarse-grained)
model if the cell size is comparable to the reaction radius, as is
rather common in physics where this is usually simply referred to
as a ``lattice model'' of reaction-diffusion \cite{RenormalizationReactionDiffusion_Review}.
But when the cells are microscopic in size, the RDME suffers from
large \emph{grid artifacts} such as dependence on the exact shape
of the cells and broken translational and rotational invariance.

An alternative approach, which removes all of the aforementioned drawbacks,
is to use a particle model of reaction-diffusion systems. The main
drawback of particle methods, which we address in this work, is their
inefficiency relative to RDME-like coarse-grained descriptions. In
particle-based models, molecules of each reactive species are tracked
explicitly, and can appear, disappear or change species via reactions.
This leads to a grid-free method that takes as input well-defined
microscopic rates. Particle-based reaction-diffusion models are a
combination of two models, a model of diffusion and a model of reaction.
A typical diffusion model employed in biochemical modeling is that
the individual particles diffuse as \emph{uncorrelated} Brownian walkers,
with each species $k$ having a specified diffusion coefficient $D_{k}$.
While this model is not easily justified in either gases or liquids,
it is commonly used and we will adopt it in this work, postponing
further discussion to the Conclusions. There are three commonly-used
reactive models, which we explain on the binary reaction $A+B\rightarrow C$
involving two reacting particles of species $A$ and $B$. Note that
reactions involving more than two particles are microscopically extremely
improbable and do not need to be considered. In the \emph{surface
reactivity} or Smoluchowski model, two particles of species $A$ and
$B$ react as soon as they approach closer than a distance $R_{AB}$,
which defines the \emph{reactive radius} for a particular reaction.
As explained in detail in the Introduction of \cite{DoiModel_Erban},
matching this kind of microscopic reaction model to measured macroscopic
reaction rates typically requires reaction radii that are too small;
using a reactive radius comparable to the expected steric exclusion
between molecules leads to a notable over-estimation of the reaction
rate. Furthermore, the Smoluchowski model is difficult to make consistent
with detailed balance (and thus all of equilibrium statistical mechanics)
for reversible reactions. This is because microscopic reversibility
requires that the reverse reaction $C\rightarrow A+B$ place the products
at a distance exactly equal to $R_{AB}$, which would lead to their
immediate reaction. One way to resolve this problem is to allow some
contacts between particles $A$ and $B$ to be non-reactive, i.e.,
to replace the fully absorbing boundary conditions in the Smoluchowski
model with partial absorption Robin boundaries, as proposed by Collins
and Kimball \cite{DiffusionControlled_RobinBC}, and studied in more
detail and employed in simulations in a number of subsequent papers
\cite{CollinsKimball_1,CollinsKimball_2,CollinsKimball_3,CollinsKimball_4}.

The \emph{volume reactivity} or Doi model \cite{ReactionDiffusion_Doi,DoiModel_Erban}
corrects the shortcomings of the surface reactivity models. Somewhat
ironically, the model was proposed by Doi in his seminal work \cite{ReactionDiffusion_Doi}
only as a way to study the Smoluchowski model in a mathematically
simpler way; recently Erban and Chapman \cite{DoiModel_Erban} suggested
that this model has a lot of merit on its own right. We employ the
Doi model in this work and therefore it is important to formulate
it precisely. In this model, all pairs of particles of species $A$
and $B$ that are closer than the reactive radius $R_{AB}$ can react
with one another as a Poisson process of rate $\lambda$, that is,
the probability that a pair of particles for which $r_{AB}\leq R_{AB}$
react during an infinitesimal time interval $dt$ is $\lambda\,dt$
\footnote{The same model applies for reactions between particles of the same
species, $A+A\rightarrow\cdot$, that is, a Poisson clock with rate
$\lambda$ ticks for each \emph{pair} of particles while they are
closer than the reactive distance. }. Since this model takes as input two rather than one parameter, one
can adjust them independently to model experimental systems more realistically.
Notably, the reactive radius can be chosen to have a realistic value
corresponding to the physical size of the reacting molecules, and
the reaction rate $\lambda$ can be tuned to reproduce measured macroscopic
rates. Nevertheless, we admit that in practical applications the Doi
model itself is a rather crude approximation of the actual molecular
structure and reaction mechanisms, and it may be difficult to define
precisely and measure accurately anything other than some effective
macroscopic reaction rates. As we explain in more detail later, achieving
microscopic reversibility (detailed balance) is quite straightforward
in the Doi model. Furthermore, while the Smoluchowski model is inherently
diffusion-limited, the Doi model can be used to study either reaction-limited
($\lambda$ small) or diffusion-limited systems ($\lambda$ large).
In the limit of infinitely fast reactions, $\lambda\rightarrow\infty$,
the Doi model becomes equivalent to the Smoluchowski model, whereas
in the limit $\lambda\rightarrow0$ one obtains a well-mixed model.

There are a number of existing particle-based algorithms for simulating
reaction-diffusion systems. For the Smoluchowski model, an exact and
efficient algorithm is the First Passage Kinetic Monte Carlo (FPKMC),
first proposed by Oppelstrup \emph{et al. }in \cite{FPKMC_PRL} and
then extended and improved in \cite{FPKMC1,FPKMC2,FPKMC_drift}, and
generalized to lattice models in \cite{FPKMC_lattice}. The event-driven
FPKMC algorithm is a combination of two key ideas: solving pair problems
analytically as in the Green's Function Reaction Dynamics (GFRD) \cite{GFRD,GFRD_KMC},
and using protective domains to ensure an exact breakdown of the multibody
problem into pairwise problems \footnote{Subsequent to its development, the FPKMC algorithm was incorporated
in the original GFRD method to yield the widely-used eGFRD method
\cite{eGFRD}.}, as in diffusion Monte Carlo methods used in quantum mechanics \cite{KLV_PRE}.
The FPKMC algorithm can, in principle, be generalized to the Doi model,
but not without sacrificing its exactness. Namely, while for the Smoluchowski
model the multibody reactive hard-sphere problem can always be broken
into two-body problems, in the Doi model three or more particles can
be within a reactive distance. However, FPKMC is not efficient at
higher density, where it becomes more and more difficult to break
the multibody problem into few-body problems. For these reasons, we
pursue a different approach in this work, which is most efficient
at higher densities and is rather straightforward to implement compared
to FPKMC.

Prior approaches to simulating the Doi model have been based on \emph{time
splitting} between diffusion and reaction. Detailed algorithmic descriptions
are missing in prior work, but the basic idea is as follows. Given
a time step size $\D t$, the particles are first diffused over the
time interval $\D t$, and then reactions are processed between pairs
of particles within the reactive radius; this is similar to what is
done in a number of popular software packages for particle modeling
of reaction-diffusion such as Smoldyn \cite{Smoldyn_v2} and MesoRD
\cite{MesoRD}. Two main factors make such particle methods inefficient.
The first one is the use of hopping in small intervals $\D t$ to
bring particles to react; this is completely bypassed by the FPKMC
method \cite{FPKMC_PRL} at the expense of algorithmic complexity.
The second factor is that at \emph{every} time step, one must do neighbor
searches in order to identify pairs of particles that \emph{may} react
during the time step. This is a necessary cost when reaction-diffusion
is combined with molecular dynamics \cite{BD_MD_ReaDDy}, but, as
we show here, is a superfluous cost when simulating simple reaction-diffusion
processes. Furthermore, even with the expensive neighbor searches
performed, the algorithm used by Robinson \emph{et al.} \cite{DoiNumerics_Erban}
employs some approximations when processing reactions, which can introduce
uncontrolled bias. For example, the reactions are processed in turn
in the same order, and it is assumed that the probability of reaction
during the time interval is small, $\lambda\,\D t\ll1$.

In this paper, we develop Split Reactive Brownian Dynamics (SRBD)
as an efficient algorithm for simulating the Doi model with \emph{controlled
accuracy}. Our algorithm still uses time splitting between diffusion
and reaction, and therefore, will be less efficient at low densities.
However, unlike existing methods, it bypasses the need to find pairs
of nearby particles at each time step. In this sense, SRBD gains efficiency
not by increasing $\D t$ as FPKMC does, but rather, by reducing the
cost of one time step. Notably, if the reaction rate is small and
reactions are infrequent, the algorithm adds a minimal cost per time
step for processing the reactions, on top of the cost of diffusing
the particles. If the reaction rate is large and a particle can undergo
more than one reaction per time step, the algorithm correctly accounts
for this. The \emph{only} error introduced is the splitting error,
and therefore the error can be controlled easily by reducing the time
step so that particles diffuse only a fraction of the reactive radius
per time step.

In Section \ref{sec:SRBD} we present the microscopic reactive model
used in SRBD, along with algorithmic details of how diffusion and
reaction are handled in the SRBD time stepping scheme. In Section
\ref{sec:Results}, we use the SRBD algorithm to study the differences
between reaction-limited and diffusion-limited regimes for irreversible
dissociation, reversible association, and a pattern-forming system.
We offer some conclusions and a discussion of open problems in Section
\ref{sec:Conclusion}.

\section{\label{sec:SRBD}Stochastic Reactive Brownian Dynamics}

In SRBD, we cover the domain with a regular grid of cells. However,
unlike the reactive cells in RDME, and like cells used in other particle
algorithms such as molecular dynamics, in SRBD the cells are only
used to help efficiently identify pairs of particles that are closer
than their reactive distance, and the size of the cells only affects
the computational efficiency but does \emph{not} affect the results.
We denote the largest reactive distance among all possible binary
reactions with $R_{\max}$. The only condition on the cell size in
SRBD is that all cells must be larger than $R_{\max}$ in each dimension.
This ensures that only particles that are in neighboring cells (we
count a given cell as a neighbor of itself) can be within a reactive
distance of one another. If the set of possible reactions changes
with time and at some point $R_{\max}$ exceeds the current cell size,
one can simply re-generates the grid of cells at the beginning of
the next time step.

The SRBD algorithm itself is a combination of the key ideas behind
two existing methods. The first one is the next subvolume method \cite{NextSubvolumeMethod_MesoRD}
for solving the RDME exactly, and the second one is the Isotropic
Direct Simulation Monte Carlo (I-DSMC) method for simulating a stochastic
hard sphere dynamics model of a fluid \cite{SHSD_PRL,SHSD}. Unlike
the next subvolume method, which treats diffusion using a master equation
as just one more reactive channel, we treat diffusion separately from
reaction. This allows one to substitute the ``motion module'' from
simple diffusion of uncorrelated walkers to something more realistic,
such as hydrodynamically correlated walkers \cite{DiffusionJSTAT}.
Following the diffusive propagation of the particles, we process reactions
between reactive pairs using an \emph{exact} event-driven algorithm
in which reactions are scheduled in an event queue and processed one
by one. As in the next subvolume method, reactive events are scheduled
\emph{per cell} rather than per particle as in FPKMC. As in the I-DSMC
method (and also the CRDME method), reactions can take place between
molecules in two disjoint neighboring cells, using \emph{rejection}
to correct for the fact that pairs of particles chosen to react may
not actually be within the reactive distance.

In this section we give a complete description of the SRBD algorithm,
as implemented in Fortran in a code that is available freely at \url{https://github.com/stochasticHydroTools/SRBD}.
We assume periodic boundary conditions throughout this paper. We will
focus the algorithmic description on the handling of binary reactions
involving two species, $A+B\rightarrow\dots$, or involving the same
species $A+A\rightarrow\dots$. However, it should be clear that the
algorithm can be trivially generalized to handle reaction networks
containing many competing reaction channels and many species (as done
in our code).

For comparison, in our code and in the results reported here, we have
also studied a method in which the diffusion model is the same as
in SRBD (independent Brownian walkers), but the reactions are performed
using a grid of cells as in the RDME. The difference with the RDME
is that the particles diffuse via a continuous random walk instead
of a jump process. This kind of model was proposed and studied in
\cite{BD_RDME}, and an algorithm was developed based on time splitting
of reaction and diffusion. We modify the method described in \cite{BD_RDME}
in order to improve translational (Galilean) invariance. Namely, before
performing reactions, we randomly shift the grid of reactive cells
by an amount uniformly distributed in $[-h/2,h/2)$, where $h$ is
the grid spacing, along each dimension. This is commonly done in a
number of other particle methods \cite{MPCD_Review} in periodic domains.
We will refer to this approach as Split Brownian Dynamics with Reaction
Master Equation (S-BD-RME). In S-BD-RME, we employ the same microscopic
reaction rules given in Section \ref{subsec:DetailedBalance} for
SRBD, replacing ``within distance $R$'' by ``in the same reactive
cell,'' and replacing ``uniformly in a sphere centered at the $A$
with radius $R$'' with ``in the same reactive cell''. This makes
all reactions in S-BD-RME local to a reactive cell, making it possible
to parallelize the algorithm straightforwardly.

Also for comparison, we will include in a number of our tests the
traditional RDME approach. We solve the RDME not by using the expensive
next reaction method, but rather, by using an efficient (easily parallelizable)
but approximate algorithm described in detail in Appendix A in \cite{FluctReactDiff}.
This algorithm is based on time splitting of diffusion and reaction
and treating diffusion using the multinomial diffusion algorithm proposed
in \cite{MultinomialDiffusion}; the only source of error in this
algorithm is the finite size of the time step size $\D t$, and we
have confirmed that reducing $\D t$ in half does not statistically
change the results reported here. One can use the S-BD-RME algorithm
to simulate the RDME simply by replacing the continuous random walks
by discrete jumping on a lattice, and not performing random grid shifts.
However, this defeats the key efficiency of the RDME over particle
methods, namely, that one does not have to track individual particles.
Instead, in the RDME one only keeps track of the total number of particles
of each species in each reactive cell. This can be a great saving
\emph{if} there are many particles per cell, but will be less efficient
than particle tracking if there are on average fewer than one molecule
per reactive cell.

\subsection{\label{subsec:MicroscopicModel}Microscopic Model}

We denote the time-dependent positions of particle $i$ with species
$s(i)$ by $\V q_{i}\left(t\right)\in\Set R^{d}$, where the dimension
$d$ is a small integer (typically 1, 2 or 3), and $i=1,\dots,N_{p}$,
where $N_{p}(t)$ is the total number of particles at time $t$. As
explained in the Introduction, in this work we use the simplest model
of diffusion: each particle is treated as a sphere with only translational
degrees of freedom and diffuses as a Brownian walker (i.e., performs
a continuous random walk) \emph{independently} of any other particles.
We assume that all particles of species $k$ diffuse with the same
coefficient $D_{k}$, $k=1,\dots,N_{s}$. As also explained in the
Introduction, we use the Doi model \cite{DoiModel_Erban} for binary
reactions: each pair of particles of species $A$ and $B$ (respectively,
$A)$ that are closer than the reactive radius $R_{AB}$ (respectively,
$R_{AA}$) can react with one another as a Poisson process of rate
$\lambda$. Different reactions $r$, $r=1,\dots,N_{r}$, have different
rates $\lambda_{r}$ (with units of inverse time) specified as input
parameters. In this work we assume an \emph{additive} hard-sphere
model,
\[
R_{AB}=R_{A}+R_{B},\quad\mbox{ and similarly,}\quad R_{AA}=2R_{A},
\]
where the radius $R_{k}$ of particles of species $k$ is an input
parameter. This can easily be relaxed, and a different reaction distance
can be specified for each binary reaction.

A proper microscopic reaction model requires complete specification
of what happens when a chemical reaction occurs. Specifically, it
requires one to specify which reactant particles change species or
disappear, and which product particles are created and \emph{where}.
We have constructed a list of microscopic reactive rules implemented
in our code by following the principle of microscopic reversibility
or \emph{detailed balance}. We explain this in some detail because
of its importance to having a reaction-diffusion model consistent
with equilibrium thermodynamics and statistical mechanics \cite{DiffusionLimited_NonDB},
which we believe to be of utmost importance. To our knowledge, few
prior works have paid close attention to this condition; for example,
the Doi model of a reversible reaction $A+B\leftrightarrow C$ proposed
in \cite{DoiModel_Reversible_Erban} is \emph{not} consistent with
detailed balance.

\subsubsection{\label{subsec:DetailedBalance}Detailed Balance}

We postulate, consistent with a long tradition of works in statistical
mechanics \cite{KeizerBook,GardinerBook,vanKampen:07}, that chemical
reactions do \emph{not} alter the state of thermodynamic equilibrium.
Instead, the state of thermodynamic equilibrium is set by chemical
potentials only. For an ideal solution/gas of non-interacting particles,
the desired equilibrium distribution is given by a product Poisson
measure, i.e., the probability of finding any given molecule (particle)
is uniform over the domain and \emph{independent} of other molecules.
The principle of detailed balance then requires that, starting from
a configuration sampled from the equilibrium state, the probability
of observing a forward reaction is equal to that of the reverse reaction.
That is, the reaction-diffusion Markov process should be time reversible
with respect to an equilibrium distribution given by a product Poisson
invariant measure \footnote{A precise mathematical statement is that the generator of the process
should be self-adjoint with respect to an inner product weighted by
the invariant measure.}.

Satisfying detailed balance with respect to such a simple equilibrium
measure is straightforward and simply requires making each reverse
reaction be the microscopic reverse of the corresponding forward reaction.
For each forward reaction, we define a specific mechanism that we
believe is physically sensible, but note that this choice is \emph{not}
unique. However, once the forward mechanism is chosen the reverse
mechanism is \emph{uniquely} determined by detailed balance. We summarize
the forward/reverse reaction rules we have used in our code in Appendix
\ref{app:ReactionRules}.

It is important to note that we do \emph{not} require that each reaction
be reversible nor do we require the existence of a thermodynamic equilibrium.
Our method can be used without difficulty to model systems that never
reach equilibrium or violate detailed balance. For example, for a
reaction $A\rightarrow B+C$, one can treat the maximum possible distance
between the product particles $R_{BC}$ as an input parameter (unbinding
radius) potentially \emph{different} from the Doi reactive radius
(binding radius) for the reverse reaction. However, by adopting microscopically
reversible reaction rules we \emph{guarantee} that if all reactions
are reversible (with the \emph{same} binding and unbinding radius),
there exists a unique state of thermodynamic equilibrium given by
a uniform product Poisson measure; furthermore, the equilibrium dynamics
will be time reversible. This makes our model fully consistent with
the fundamental principles of equilibrium statistical mechanics; for
example, as we show in Section \ref{subsec:TailsABC}, the equilibrium
constants (concentrations) will be independent of kinetics (rates).

\subsubsection{\label{subsec:ExtraDiffusion}Enhancement of diffusion coefficients
by reaction}

It is important here to point out a somewhat unphysical consequence
of the microscopic Doi reaction model for reversible reactions: increased
species diffusion due to reactions. For specificity, let us consider
the reversible reaction $A+B\leftrightarrow C$ and denote the number
densities of the reactants with $\V n=\left(n_{A},n_{B},n_{C}\right)$.
In the RDME model, reacting particles have effectively the same position,
since they need to be on the same lattice site to react. Similarly,
the reverse reaction creates particles at the same site. This implies
that if one constructs, for example, the linear combination $\tilde{n}=n_{A}+n_{B}+2n_{C}$,
the reaction strictly preserves $\tilde{n}$ locally. This means that
$\tilde{n}$ evolves by diffusion only.

In the SRBD or S-BD-RME models, however, the positions of the reacting
particles are off lattice. This means that, even in the absence of
explicit particle diffusion, the quantity $\tilde{n}$ will evolve
due to reactions. In particular, assume that an $A$ and a $B$ particle
react and the $B$ becomes a $C$, and thereafter the $C$ decays
via the reverse reaction into a $B$ and creates an $A$ within a
reactive radius of the original $B$. This net consequence of this
sequence of successive forward and reverse reactions is that the $A$
particle has randomly displaced, from one position inside a reactive
sphere centered at the $B$ particle, to a new position within this
sphere. This is an effective diffusive jump of the $A$ particle and
leads to an enhancement of the diffusion coefficient of the quantity
$\tilde{n}$ on the order of $R_{AB}^{2}/\tau$, where $\tau$ is
a reactive time for the sequence $A+B\rightarrow C\rightarrow A+B$.
This effective enhancement can become important when $R_{AB}$ is
large, and lead to different \emph{physical} behavior between the
RDME and the SRBD or S-BD-RME models, as we will demonstrate explicitly
in Sections \ref{subsec:TailsABC} and \ref{subsec:TuringPatterns},
and discuss further in the Conclusions.

\subsection{Time Splitting}

The \emph{only} approximation we make in our SRBD algorithm is that
diffusion and reaction are handled separately by using time splitting.
We use the well-known second-order Strang splitting method, and then
solve the diffusion and reaction subproblems \emph{exactly}. Specifically,
one time step of duration $\D t$ going from time level $n$ to $n+1$
is performed as described in Algorithm \ref{alg:SplitStep}. The errors
induced in expectation values (observables) by the Strang splitting
are of order $O\left(\D t^{2}\right)$.

\begin{algorithm}
\begin{enumerate}
\item Diffuse all particles $i$ present at time $t^{n}$ for half a time
step
\[
\V q_{i}^{n+\frac{1}{2}}=\V q_{i}^{n}+\sqrt{D_{s(i)}\D t}\,\V{\mathcal{N}}\left(0,1\right),
\]
where $\V{\mathcal{N}}\left(0,1\right)$ denotes a Gaussian random
variate of mean zero and unit variance, sampled independently for
each particle and at each time step.
\item \label{enu:ReactionStep}Process reactions exactly over a time interval
$\D t$ during which the particles do not diffuse (i.e., they stay
in place) using Algorithm \ref{alg:ReactionStep}. If using S-BD-RME,
choose a grid of cells to perform the reactions in, potentially randomly
shifted from a reference grid to improve translational invariance,
and then process reactions in each cell independently using the traditional
SSA algorithm.
\item Diffuse all remaining and newly-created particles $i$ for half a
time step
\[
\V q_{i}^{n+1}=\V q_{i}^{n+\frac{1}{2}}+\sqrt{D_{s(i)}\D t}\,\V{\mathcal{N}}\left(0,1\right).
\]
\end{enumerate}
\caption{\label{alg:SplitStep}Split time step of the SRBD or S-BD-RME algorithms.}
\end{algorithm}

\subsection{Processing Reactions in SRBD}

We now turn our attention to the core of the SRBD algorithm, processing
reactions over a time interval $\D t$ during which the particles
do not diffuse. We perform this reaction-only step exactly, i.e.,
we sample from the correct Markov process without any approximations.
We focus here on processing binary reactions since reactions involving
zero or one reactants are trivial to handle.

A naive algorithm would proceed as follows. First, for each binary
reaction, create a list of \emph{all} pairs of particles that \emph{could}
react because they are within a reactive distance of each other (note
that one pair can in principle participate in a number of different
reactions). Let the number of potential pairs for reaction $r$ be
$N_{pr}$. Set the propensity (rate) for reaction $r$ to $N_{pr}\lambda_{r}$,
and select the next reaction to happen and increment the time counter
(if less than $\D t$) using the traditional SSA algorithm. Once a
specific reaction is selected, select one of the $N_{pr}$ pairs for
that reaction uniformly at random, and process the reaction using
the microscopic reaction rules described in Appendix \ref{app:ReactionRules}.
Update the list of pairs and repeat the process until a time $\D t$
has elapsed. This algorithm, while clearly correct, is very inefficient
due to the need to search for pairs of overlapping particles at each
step, and to update this list after each reaction is processed. In
Algorithm \ref{alg:ReactionStep} we summarize an algorithm that is
identical in law. In other words, the algorithm selects pairs of particles
with the correct probability rates, \emph{without} ever explicitly
searching for pairs within a reactive distance. We detail the various
steps in subsequent sections.

Note that the algorithm presented here requires having efficient \emph{dynamic}
spatial data structures that enable finding all particles that are
in a given cell, as well as keeping an accurate count of the number
of particles of each species in each cell. The most efficient, in
both storage and time, and simplest data structure that can be used
for this purpose is what we will refer to as linked-list cells (LLCs)
\cite{AED_Review}. These are essentially integer-based \emph{linked
lists}, one list for each cell and each species, that stores an integer
identifier for all particles of the given species in the given cell,
see the book \cite{Allen_Tildesley_book} for implementation details.
Additionally, one requires a data structure for managing the \emph{event
queue}; our implementation uses a \emph{heap} of fixed size equal
to the number of cells for this purpose.

\begin{algorithm}
\begin{enumerate}
\item Prepare: Build linked-list cells (LLCs) and reset the event queue.
\item \label{enu:Schedule1}Sample the time to the next reaction for each
cell $i$ (see Algorithm \ref{alg:SchedulingReactions}) $\d t_{i}$
and compute the scheduled next reaction time $t_{i}=t+\d t_{i}$.
If the scheduled time $t_{i}\leq t^{n}+\D t$, insert the scheduled
reaction event into the event queue with time stamp $t_{i}$.
\item Event loop: Until the event queue is empty, do:

\begin{enumerate}
\item Select cell $i$ on top of the queue with time stamp $t_{i}$, $t^{n}\leq t_{i}\leq t^{n}+\D t$,
and advance the global time to $t=t_{i}$.
\item Select next reaction to happen in cell $i$ using a traditional KMC/SSA
method.
\item If the reaction is unary, select a particle in cell $i$ to undergo
the chosen reaction. If the reaction is binary, select a pair of particles,
one in cell $i$ and the other in one of its neighboring cells, to
undergo the chosen reaction. For binary reactions, if particles do
not overlap, i.e., they are not within a reactive radius of each other,
then skip to step \ref{enu:Schedule2}.
\item \label{enu:Process}Process the reaction (see Algorithm \ref{alg:Processing}),
creating/destroying particles and updating the LLCs as necessary.
\item \label{enu:Schedule2}For each cell $j$ that is a neighbor of cell
$i$ (including cell $i$) that was (potentially) affected by the
reaction, reschedule the time to the next reaction $\d t_{j}$ (see
Algorithm \ref{alg:SchedulingReactions}). If $t_{j}=t+\d t_{j}<t+\D t$
schedule the next event for cell $j$ at time $t_{j}$ and update
the event queue, otherwise delete cell $j$ from the queue.
\end{enumerate}
\end{enumerate}
\caption{\label{alg:ReactionStep}Summary of the SRBD reaction step \ref{enu:ReactionStep}
in Algorithm \ref{alg:SplitStep} during the $n$-th time step.}
\end{algorithm}

\subsubsection{\label{subsec:SchedulingReactions}Scheduling reactions}

For unary reactions such as decay $A\rightarrow\dots$, what we mean
by a reaction occurring in cell $i$ is that the particle undergoing
the reaction is in cell $i$. For binary hetero-reactions with different
reactants $A+B\rightarrow...$ with Doi rate $\lambda$, we schedule
separately the two ordered reactions $A+B\rightarrow...$ and $B+A\rightarrow...$,
i.e., we distinguish one of the two particles as the ``first'' particle
and the other as the ``second'' particle. Each of these two reactions
occurs with half the rate of the original (unordered) reaction. We
associate a given binary reaction to the cell in which the first particle
is located. For binary reactions with a single species, $A+A\rightarrow\dots$,
we also order the two reacting particles into a first and a second
particle, and associate the reaction with the cell of the first particle.
The effective rate of the ordered reactions is again half of the original
rate because each pair is scheduled for reaction twice. A little thought
reveals that no special treatment is needed for a homo-reaction $A+A\rightarrow\dots$,
where both of the particles are in the same cell except for rejecting
self-reactions; each pair of particles can again be selected twice.

Scheduling the time $\d t$ until the next reaction to occur in a
given cell is done by computing the total reaction rate (called propensity
in the SSA literature) $\alpha$ over all reactions associated with
that cell, and sampling an exponentially-distributed time lag $\d t$
with mean $\alpha^{-1}$, as detailed in Algorithm \ref{alg:SchedulingReactions}.
Note that the computation in Algorithm \ref{alg:SchedulingReactions}
over-estimates the actual rate for binary reactions since it does
not account for whether the particles actually overlap; we correct
for this using rejection. Specifically,\textcolor{red}{{} }if a pair
selected to react does not overlap, we reject the pair. Similarly,
for homo-reactions $A+A\rightarrow\dots$ we correct for the fact
that we compute the number of pairs of particles of species $A$ as
$N_{A}^{2}/2$ instead of $N_{A}\left(N_{A}-1\right)/2$ by rejecting
reactions of a particle with itself.

\begin{algorithm}
\begin{itemize}
\item For each binary reaction $r$ for cell $i$, $A+B\rightarrow...$
where we allow for the possibility that $B=A$ and the order of the
reactants matters if $B\neq A$, compute the rate in cell $i$ as
\[
\alpha_{r}=\frac{\lambda}{2}N_{A}N_{B}^{\prime}
\]
where $N_{A}$ is the number of $A$ particles in cell $i$, and $N_{B}^{\prime}$
is the total number of $B$ particles in the neighborhood of $i$.
\item Add the rates of all possible reactions, $\alpha=\sum_{r=1}^{N_{r}}\alpha_{r}$
(as in ordinary SSA).
\item Sample an exponentially distributed random number $\d t_{i}$ with
mean $\alpha^{-1}$.
\end{itemize}
\caption{\label{alg:SchedulingReactions}Algorithm used to schedule the time
to the next reaction $\protect\d t_{i}$ for cell $i$ during steps
\ref{enu:Schedule1} and \ref{enu:Schedule2} in Algorithm \ref{alg:ReactionStep}.}
\end{algorithm}

\subsubsection{\label{subsec:ProcessingReactions}Processing binary reactions}

Once a given cell is chosen to have a reaction occur in it, one must
select one (for unary reactions) or two (for binary reactions) particles
to participate in the reaction, randomly and uniformly from among
all particles of the required species that are in the given cell or
one of its neighboring cells. This step is made efficient by using
LLCs and the counts of the number of particles of each species in
each cell. We give additional details of the processing of binary
reactions in Algorithm \ref{alg:Processing}.

\begin{algorithm}
\begin{itemize}
\item Randomly and uniformly select a particle of species $A$ that is in
cell $i$, and another particle of species $B$ from a cell $j$ that
neighbors cell $i$. Note that this can select the same particle twice
if $B=A$; when this happens, skip the remaining steps.
\item Test if the two particles are within their reactive distance, and
if not, skip the remaining steps.
\item Otherwise, process the reaction by deleting and adding particles depending
on the reaction products, following the microscopic reaction rules
explained in Appendix \ref{app:ReactionRules}. While processing the
reactions, keep the LLCs up to date and keep track of whether any
reaction changes the population of cell $i$ (number of particles
of each species in that cell), and also whether the population of
cell $j$ changes.
\item Reschedule the time to the next reaction $\d t_{i}$ for cell $i$
using Algorithm \ref{alg:SchedulingReactions}. If $t_{i}=t+\d t_{i}<t+\D t$
schedule the next event for cell $i$ at time $t_{i}$ and update
the event queue, otherwise delete cell $i$ from the queue.
\item If the population of cell $i$ changed, update the event prediction
for all neighbor cells of $i$.
\item If the population of cell $j$ changed, update the event prediction
for all cell neighbors of $j$ that are not also neighbors of $i$.
\end{itemize}
\caption{\label{alg:Processing}Algorithm used in steps \ref{enu:Process}
and \ref{enu:Schedule2} of Algorithm \ref{alg:ReactionStep} to process
a binary reaction $A+B\rightarrow\dots$ (where it may be that $B=A$)
associated with cell $i$ at global time $t$.}
\end{algorithm}

\section{\label{sec:Results}Examples}

In this section we apply the SRBD algorithm to a number of reaction-diffusion
problems, and compare the numerical results to theoretical predictions
and results obtained using RDME, as well as S-BD-RME. First, in Section
\ref{subsec:MicroToMacro}, we explore the relationship between microscopic
reaction rates and the effective macroscopic rates for reaction- and
especially diffusion-limited irreversible bimolecular reactions in
three dimensions. In Section \ref{subsec:OptimalGrid} we briefly
discuss how the choice of the grid cell size in the SRBD algorithm
affects the computational efficiency of the algorithm. In Section
\ref{subsec:TailsABC} we study the dynamics of concentration fluctuations
at thermodynamic equilibrium for diffusion-limited reversible association
in two dimensions. Lastly, in Section \ref{subsec:TuringPatterns}
we study the formation of Turing-like patterns in a reaction-limited
two-dimensional system.

\subsection{\label{subsec:MicroToMacro}Conversion from microscopic to macroscopic
reaction rates}

One of the key issues when comparing different methods, such as SRBD
and RDME, is ensuring that the parameters in the two models are both
consistent with the same effective \emph{macroscopic} model at length
scales much larger than the discretization scale (i.e., reactive distances
in SRBD or grid size in RDME), and time scales much larger than the
microscopic ones. This is especially difficult to do because the effective
macroscopic model is non-trivial to obtain for reaction-diffusion
problems, and is \emph{not} always given by a simple deterministic
reaction-diffusion partial differential equation. In particular, a
large body of literature has emerged over the past several decades
showing that the macroscopic behavior is unusual for \emph{diffusion-limited
}systems, i.e., systems in which reactions happen quickly once reactants
find each other in physical proximity \cite{AsymptoticPowerLaw_ABC,LongTimeTails_ABC,RelaxationTime_ABC,Coagulation_Renormalization,Coagulation_Renormalization_PRL,DiffusionLimited_NonDB,DiffusionLimitedAnnihilation}.
For example, the traditional law of mass action is known to break
down in simple diffusion-limited coagulation even in three dimensions
\cite{Coagulation_Renormalization,Coagulation_Renormalization_PRL},
making it impossible to even define what is meant by an effective
or macroscopic reaction rate. Even when the law of mass action is
formally recovered for the \emph{instantaneous }reaction rate, long-term
memory effects appear in the long-time dynamics \cite{AsymptoticPowerLaw_ABC,LongTimeTails_ABC}.
In general, in diffusion-limited systems nontrivial correlations between
fluctuations of the number densities of different species appear at
molecular scales, and diffusion coefficients enter in the macroscopic
``reaction rates'' in addition to the microscopic reaction rates.
This is to be contrasted with the much simpler behavior for \emph{reaction-limited}
systems, where diffusion dominates and uniformly mixes the reactants,
thus eliminating microscopic spatial correlations between different
species.

To illustrate these subtle physical effects and confirm that the SRBD
model and algorithm produce the correct results both in reaction-limited
and diffusion-limited settings, we study here two simple examples
for which analytical predictions are available. The first is a one-species
model of coagulation, $A+A\rightarrow A$, and the second one is a
two-species model of annihilation, $A+B\rightarrow B$, both of which
we study in three dimensions. We define an effective macroscopic binary
reaction rate $k$ as follows. We insert particles of species $A$
randomly and uniformly into the system via the reaction $\emptyset\underset{k_{i}}{\rightarrow}A$,
and wait until a steady state is established at a given average (over
space and time) number density $n_{A}$. The effective forward rate
for the binary reaction at the steady state number density (observe
that the average number density of $B$ molecules is unchanged by
the reaction $A+B\rightarrow B$) is then defined as $k=k_{i}/\left(n_{A}n_{B}\right)$
for $A+B\rightarrow B$, and $k=k_{i}/n_{A}^{2}$ for $A+A\rightarrow A$.

\subsubsection{Reaction-Limited versus Diffusion-Limited Rates}

For sufficiently low packing densities $\phi$ (defined precisely
later), one can estimate the effective (macroscopic) association or
forward reaction rate $k_{0}=\lim_{\phi\rightarrow0^{+}}k$ for binary
reactions in SRBD by generalizing the approach originally proposed
by Smoluchowski for the Doi reactivity model. In this approach, many-body
effects are neglected, as later justified by Doi \cite{ReactionDiffusion_Doi}.
Details can be found in the work of Erban and Chapman \cite{DoiModel_Erban};
here we summarize the important results. For a single reaction $A+B\rightarrow\dots$
or $A+A\rightarrow\dots$ in three dimensions, at low densities, the
macroscopic reaction rate $k_{0}$ (units $\text{m}^{3}/\text{s}$)
is predicted to be related to the microscopic rate $\lambda$ (units
$\text{s}^{-1})$ and the reactive radius $R$ in the Doi model via
\cite{DoiModel_Erban}
\begin{equation}
k_{0}=s\left(4\pi DR\right)\left[1-\sqrt{\frac{D}{\lambda R^{2}}}\;\tanh\left(\sqrt{\frac{\lambda R^{2}}{D}}\right)\right],\label{eq:lambda_to_k_AB}
\end{equation}
where $s=1$, $D=D_{AB}=D_{A}+D_{B}$ and $R=R_{AB}=R_{A}+R_{B}$
for $A+B\rightarrow\dots$, and $s=1/2$, $D=2D_{A}$ and $R=R_{AA}=2R_{A}$
for $A+A\rightarrow\dots$.

Let us define a dimensionless number
\[
r=\frac{\lambda R^{2}}{D},
\]
which compares the reaction rate to the diffusion rate. First, note
that for all values of $r$ we have $k_{0}<4\pi DR$. If $r\gg1$,
the system is diffusion-limited, and $k_{0}\approx k_{S}^{\text{SRBD}}$
approaches the Smoluchowski rate $k_{S}^{\text{SRBD}}=4\pi DR$, i.e.,
the rate that would be obtained if particles reacted upon first touching.
For $r\ll1$, the system is reaction-limited, and we obtain
\begin{equation}
k_{0}\approx k_{\text{mix}}=s\frac{4\pi}{3}R^{3}\lambda.\label{eq:lambda_to_k_mix_3D}
\end{equation}
This result has a very simple physical interpretation. In the limit
$r\ll1$, the particle positions are uncorrelated, i.e., the system
is ``uniformly mixed'' at microscopic scales. Therefore, to find
the instantaneous reaction rate one can simply multiply by $\lambda$
the total number of overlapping particle pairs $V_{r}n_{A}n_{B}$
(taking $n_{B}\equiv n_{A}$ for the one-species case), where $V_{r}=4\pi R^{3}/3$
is the reactive volume, which gives the total reaction rate as $k_{0}n_{A}n_{B}\approx\lambda V_{r}n_{A}n_{B}.$
The formula (\ref{eq:lambda_to_k_mix_3D}), unlike (\ref{eq:lambda_to_k_AB}),
applies at \emph{all densities}, i.e., $k\approx k_{0}$ independent
of the density for reaction-limited systems. Furthermore, while it
is not possible to generalize (\ref{eq:lambda_to_k_AB}) to two dimensions,
there is no problem in generalizing (\ref{eq:lambda_to_k_mix_3D})
to any dimension, simply by using the corresponding formula for the
volume of the reactive sphere $V_{r}$. Lastly, while (\ref{eq:lambda_to_k_AB})
does not generalize to the case of many species and reactions, (\ref{eq:lambda_to_k_mix_3D})
continues to apply for each reaction in reaction-limited systems.
This emphasizes the fact that reaction-limited systems are much simpler
to model than diffusion-limited ones. We will use this in Section
\ref{subsec:TuringPatterns} when studying Turing-like pattern formation
in a reaction-limited system.

One can argue that, in general, if a system is well described by (deterministic
or fluctuating) hydrodynamics, it should be in the reaction-limited
regime unless the reactants are very dilute. For a generic binary
reaction $A+B\rightarrow\dots$, an important characteristic length
scale is the so-called \emph{penetration depth}, i.e., the typical
distance a molecule travels between successive reactions, 
\[
L_{p}=\sqrt{\frac{D_{AB}}{kn_{AB}}},\quad\mbox{where}\quad n_{AB}=n_{A}+n_{B}.
\]
This length scale should be macroscopic, which means that the number
of molecules in a penetration volume $N_{L}=n_{AB}L_{p}^{3}\gg1$;
otherwise a hydrodynamic-level description would not be appropriate
on length scales of order $L_{p}$. Let us also define the packing
fraction $\phi=\left(4\pi/3\right)n_{AB}R_{AB}^{3}.$ Under this condition
we see that unless $\phi\ll1$,
\[
\frac{k}{R_{AB}D_{AB}}=\frac{1}{N_{L}}\left(\frac{L_{p}}{R_{AB}}\right)\sim\left(\phi N_{L}^{2}\right)^{-\frac{1}{3}}\ll1,
\]
which implies that the reaction is reaction-limited, $r\ll1$. Before
we return to a reaction-limited system in Section \ref{subsec:TuringPatterns},
we focus on the harder case of diffusion-limited systems, in which
a particle-level description is required to capture the nontrivial
spatial correlations among the reactants.

\subsubsection{Diffusion-Limited Reactions}

Erban and Chapman \cite{DoiModel_Erban} have computed the conversion
from microscopic to macroscopic rates for RDME; the same formulas
have also been computed by Winkler and Frey in \cite{Coagulation_Renormalization_PRL,Coagulation_Renormalization}.
Specifically, in three dimensions the effective macroscopic rate is
related to the input (microscopic) rate $k_{\text{RDME}}$ that enters
in the RDME and the grid spacing $h$ via \cite{DoiModel_Erban,Coagulation_Renormalization_PRL}
\begin{equation}
\frac{1}{k_{0}}=\frac{1}{k_{\text{RDME}}}+\frac{\beta=0.25273}{hD},\label{eq:k_RDME_to_k}
\end{equation}
where $D=D_{AB}$ for $A+B$ and $D=D_{A}$ for $A+A$. This formula
contains the same physics as (\ref{eq:lambda_to_k_AB}), with $h$
playing the role of $R_{AB}$, and $k_{\text{RDME}}/h^{3}$ playing
the role of $\lambda$. The appropriate definition of the dimensionless
number is now $r=k_{\text{RDME}}/\left(hD\right)$. For reaction-limited
systems, $r\ll1$, the effective rate $k\approx k_{0}\approx k_{\text{RDME}}$
is the same as the microscopic rate. For diffusion-limited systems,
$r\gg1$, we have $k_{0}\approx k_{S}^{\text{RDME}}=hD/\beta$, which
is the RDME equivalent of the Smoluchowski/Doi formula $k_{S}^{\text{SRBD}}=4\pi DR$,
and no longer involves the precise value of $k_{\text{RDME}}$. If
one keeps $k_{\text{RDME}}$ and $D$ fixed but reduces the grid spacing
$h$, one gets $k_{0}\rightarrow0$, i.e., binary reactions are lost
because reactants can no longer find each other by diffusion \cite{MesoRD_GridResolution,RDME_Bimolecular_Petzold,CRDME}.

The simple analytical results (\ref{eq:lambda_to_k_AB}) and (\ref{eq:k_RDME_to_k})
are limited to low densities since they neglect all many-body effects.
The only theory we are aware of for non-vanishing densities is the
renormalization group analysis of Winkler and Frey \cite{Coagulation_Renormalization_PRL,Coagulation_Renormalization}
for the coagulation reaction $A+A\rightarrow A$ in three dimensions.
The nontrivial computation detailed in \cite{Coagulation_Renormalization}
predicts that the leading order correction to (\ref{eq:k_RDME_to_k})
is given by the \emph{non-analytic} correction (see (33) in \cite{Coagulation_Renormalization})
\begin{equation}
\frac{k}{k_{0}}=1+\alpha\left(\frac{k_{0}}{D_{A}}\right)^{\frac{3}{2}}n_{A}^{\frac{1}{2}}=1+\gamma\left(\frac{k_{0}}{k_{S}^{\text{RDME}}}\right)^{3/2}\phi^{\frac{1}{2}},\label{eq:k_renormalized_coagulation}
\end{equation}
where $k_{0}=\lim_{\phi\rightarrow0}k$ is given by (\ref{eq:k_RDME_to_k}),
$n_{A}$ is the number density of $A$ molecules with $\phi=n_{A}h^{3}$
being the packing density, $\alpha=\left(2\pi\sqrt{2}\right)^{-1}$
is a universal constant for some set of models that are invariant
under the renormalization group \cite{Coagulation_Renormalization_PRL},
and $\gamma=\alpha/\beta^{3/2}$ is a constant. We expect a similar
formula to apply to SRBD as well, defining the packing fraction as
$\phi=n_{A}\cdot\left(4\pi R_{AA}^{3}/3\right)$; however, the renormalization
group analysis performed in \cite{Coagulation_Renormalization_PRL}
does not apply to the Doi reactivity model, and a different coefficient
$\gamma$ is expected. We are not aware of any finite-density theory
for a reaction involving two different species.

In order to confirm that our SRBD algorithm correctly reproduces these
rigorous theoretical results, we perform simulations of diffusion-limited
coagulation and annihilation reactions. First, we consider a system
with only one species and two reactions, $A+A\rightarrow A$ and $\emptyset\underset{k_{i}}{\rightarrow}A$.
We have confirmed that the results presented here are not affected
by finite-size artifacts by increasing the system size (i.e., the
number of particles) up to as many as $128^{3}$ cells (for both SRBD
and RDME, where in SRBD we set the cell size to $h=R$) \footnote{The Smoluchowski theory for low dilution can easily be modified to
have a finite cutoff, and leads to an estimate for the relative correction
due to finite system size of order $\sim O(1)/\left(L/R\right)$,
where $L$ is the system size.}. In arbitrary units, the SRBD parameters chosen are $R=R_{AA}=2R_{A}=1$
and $\lambda=0.4775$ (corresponding to $k_{\text{mix}}=1.0$), and
$D_{A}=0.03255$, giving $r\approx7.33$ and therefore $f=k_{0}/k_{S}^{\text{SRBD}}=0.6340$.
For the RDME, we set $h=1$, $D_{A}=0.25272$ and set the microscopic
binary reaction rate to $k_{\text{RDME}}=1$, giving $f=k_{0}/k_{S}^{\text{RDME}}=0.5$.

\begin{figure}
\begin{centering}
\includegraphics[width=0.49\textwidth]{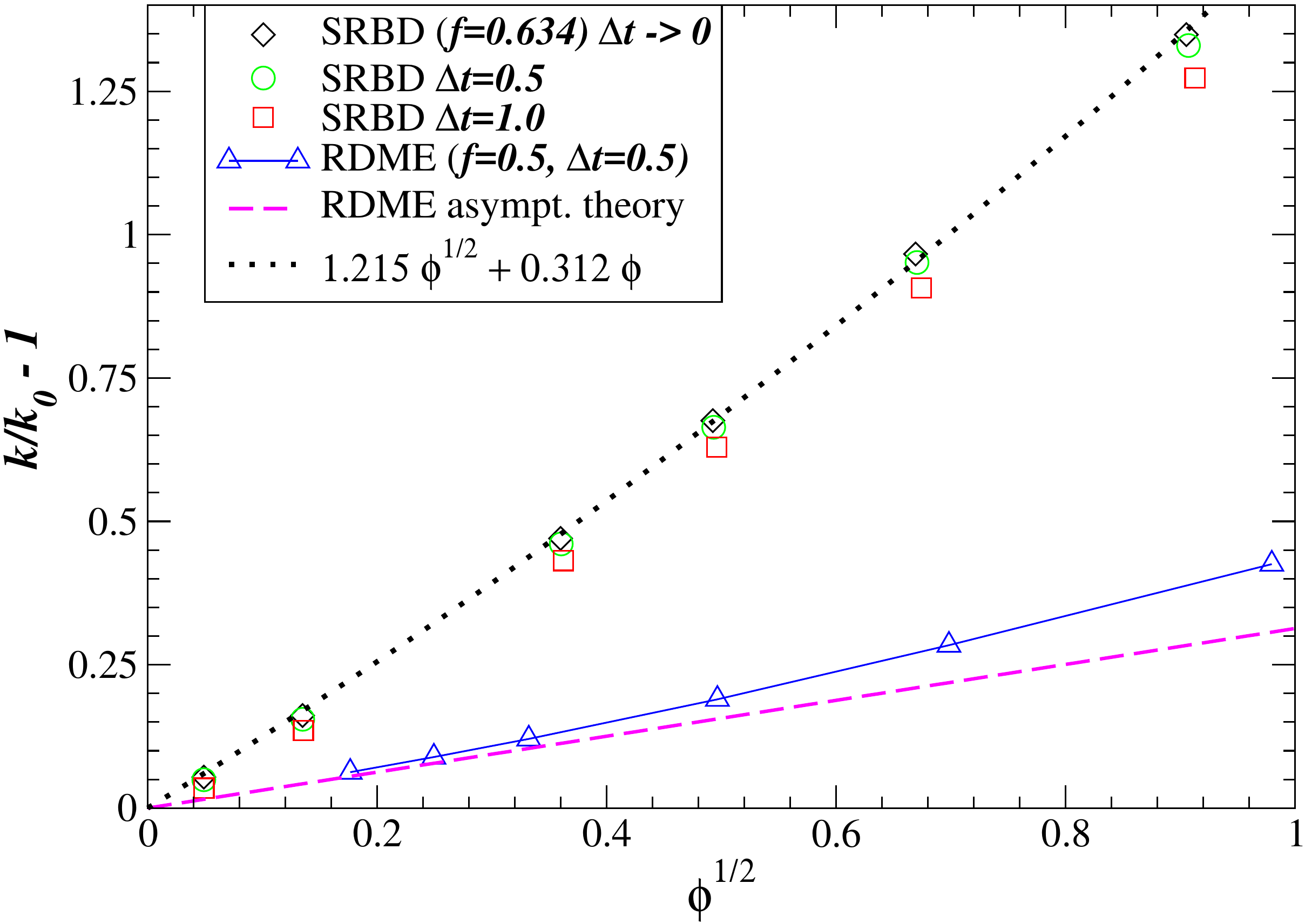}\includegraphics[width=0.49\textwidth]{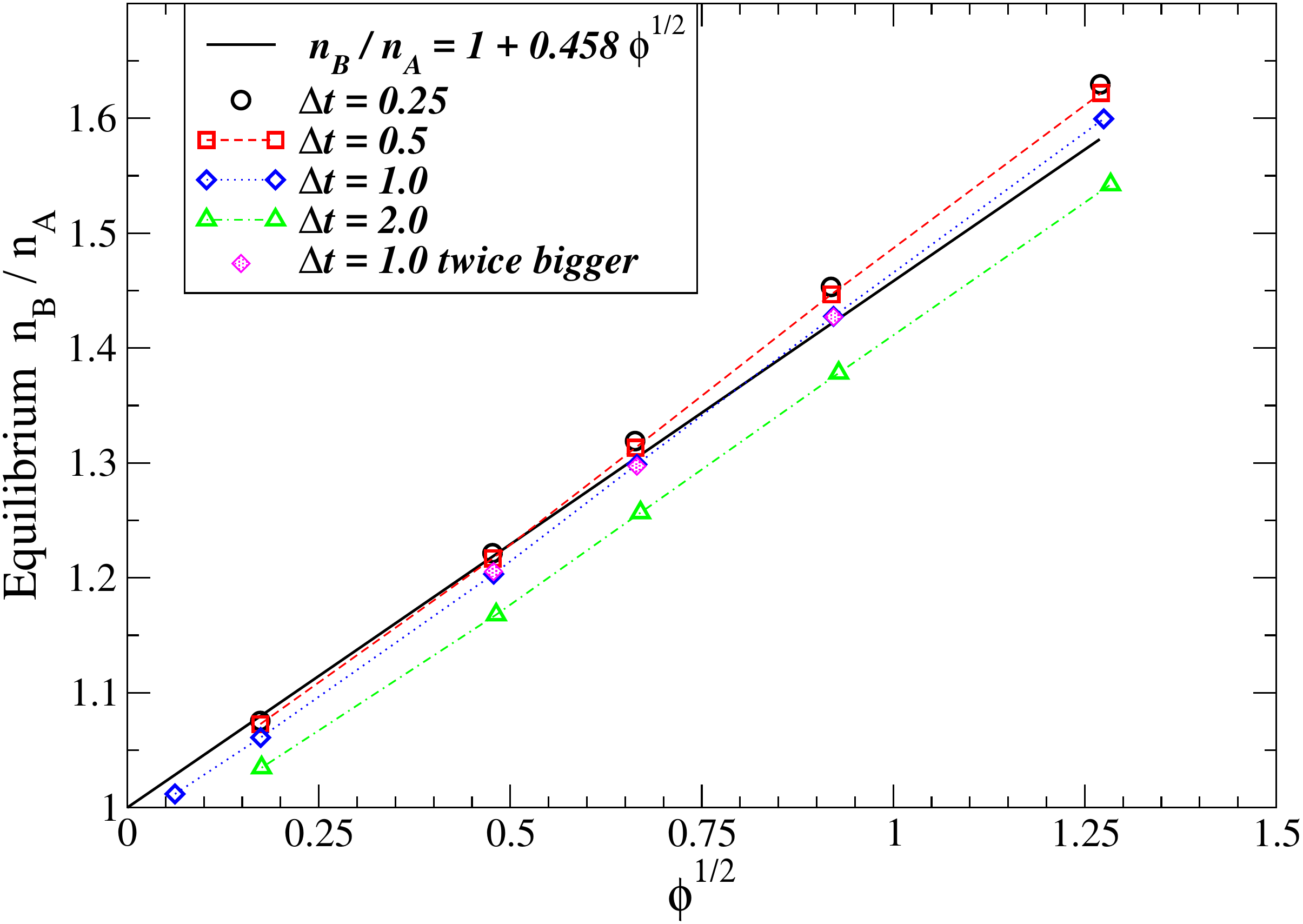}
\par\end{centering}
\caption{\label{fig:MicroToMacro}Conversion from microscopic to macroscopic
reaction rates for (partially) diffusion-limited systems ($f=k_{0}/k_{S}\sim1/2$)
at finite packing densities $\phi$. Error bars are smaller or comparable
to symbol size. (\emph{Left}) Correction to the low-density rate $k_{0}$,
given by (\ref{eq:lambda_to_k_AB}) for SRBD and by (\ref{eq:k_RDME_to_k})
for RDME, for coagulation, $A+A\rightarrow A$ and $\emptyset\rightarrow A$.
For SRBD, we use two different time step sizes (see legend) and extrapolate
to the exact result without splitting errors. For the RDME an exact
renormalization calculation gives the leading order non-analytic $\phi^{1/2}$
correction \cite{Coagulation_Renormalization_PRL,Coagulation_Renormalization},
which matches our numerical results for sufficiently small densities.
For SRBD the result is well-fit by the empirical fit $k/k_{0}=1+1.215\phi^{1/2}+0.312\phi$
(dotted black line). (\emph{Right}) Deviation $k/k_{0}=n_{B}/n_{A}$
from the low-density rate $k_{0}$ given by (\ref{eq:lambda_to_k_AB})
for SRBD for annihilation, $A+B\rightarrow B$ and $\emptyset\rightarrow A$,
for several time step sizes (see legend). For $\protect\D t=1$ we
show results obtained using a system that is twice larger (i.e., eight
times the number of particles) and see no measurable finite-size effects.
There is no theory for finite densities but the result is consistent
with the empirical fit $k/k_{0}\approx1+0.458\phi^{1/2}$.}
\end{figure}

Our results for $k/k_{0}-1$ as a function of $\phi$ are shown in
the left panel of Fig. \ref{fig:MicroToMacro} for both SRBD and RDME.
Based on our numerical confirmation (not shown) that the temporal
discretization errors in SRBD are quadratic in $\D t$, we also show
estimated result for $\D t\rightarrow0$ obtained by extrapolating
the numerical results for two different time step sizes $\D t$. For
the RDME runs, we use $\D t=0.5$. The results demonstrate that the
SRBD results are in agreement with the known theoretical results.
In particular, we see that the extrapolated curve would pass through
the origin, indicating that $k\rightarrow k_{0}$ as $\phi\rightarrow0$,
as it must. The results for SRBD are consistent with a non-analytic
$\phi^{1/2}$ dependence for small $\phi$, but it is difficult to
say anything more quantitative due to the lack of a theoretical prediction
for the dependence on density. For RDME, we see agreement with the
theory of Winkler and Frey \cite{Coagulation_Renormalization_PRL}
for small but finite densities, although it is clear that higher-order
terms are non-negligible for $\phi\gtrsim0.5$.

For annihilation we consider a system with two species and the reactions
$A+B\rightarrow B$ (i.e., conserved number of $B$ molecules) and
$\emptyset\underset{k_{i}}{\rightarrow}A$. We change the number density
of B molecules $n_{B}$ and wait until a steady state is reached at
a specific average density $n_{A}$ of $A$ molecules. In this case
we define a packing density as $\phi=\left(n_{A}+n_{B}\right)\cdot\left(4\pi R_{AB}^{3}/3\right)$.
We set $R_{AB}=1$ and $\lambda=0.2387$ (giving $k_{\text{mix}}=1$),
and $D_{A}=D_{B}=0.03255$, giving $r\approx3.67$ and therefore $f=k_{0}/k_{S}^{\text{SRBD}}=0.5$.
We set the rate of (random) insertion $k_{i}$ of $A$'s into the
system to $k_{i}=k_{0}n_{B}^{2}$, so that if $k=k_{0}$ it would
be that $n_{A}=n_{B}$ at equilibrium. Therefore, $kn_{A}n_{B}=k_{0}n_{B}^{2}$,
giving $k/k_{0}=n_{B}/n_{A}$. In the right panel of Fig. \ref{fig:MicroToMacro}
we compute the deviation of the measured reaction rate from the low-density
prediction (\ref{eq:lambda_to_k_AB}) via the ratio $n_{B}/n_{A}$
at several packing densities and for several values of the time step
size. Again the results are consistent with a splitting error of order
$O\left(\D t^{2}\right)$ (not shown), and for sufficiently small
$\D t$ and $\phi\rightarrow0$ the results are in perfect agreement
with the theory (\ref{eq:lambda_to_k_AB}). For finite densities,
there is no known theory but we expect that for fixed composition
(\ref{eq:k_renormalized_coagulation}) will hold with a different
coefficient $\gamma$; indeed, the results appear to be consistent
with a non-analytic $\phi^{1/2}$ dependence for small $\phi$.

\subsection{\label{subsec:OptimalGrid}Optimal Cell Size}

We recall that, as far as accuracy is concerned, the choice of the
grid cell size $h$ in SRBD is arbitrary beyond requiring $h\geq R_{\max}$.
However, the choice of $h$ is crucial to the efficiency of the algorithm.
We expect that there will be an optimal choice $h_{\text{opt}}$ that
balances the increased costs of maintaining and building the grid
data structures versus the benefits of reducing the number of particle
pairs that need to be checked for overlap. If $h_{\text{opt}}<R_{\max}$
then we must set $h=R_{\max}$ in order to ensure correctness of the
algorithm.

\begin{figure}
\begin{centering}
\includegraphics[width=0.7\textwidth]{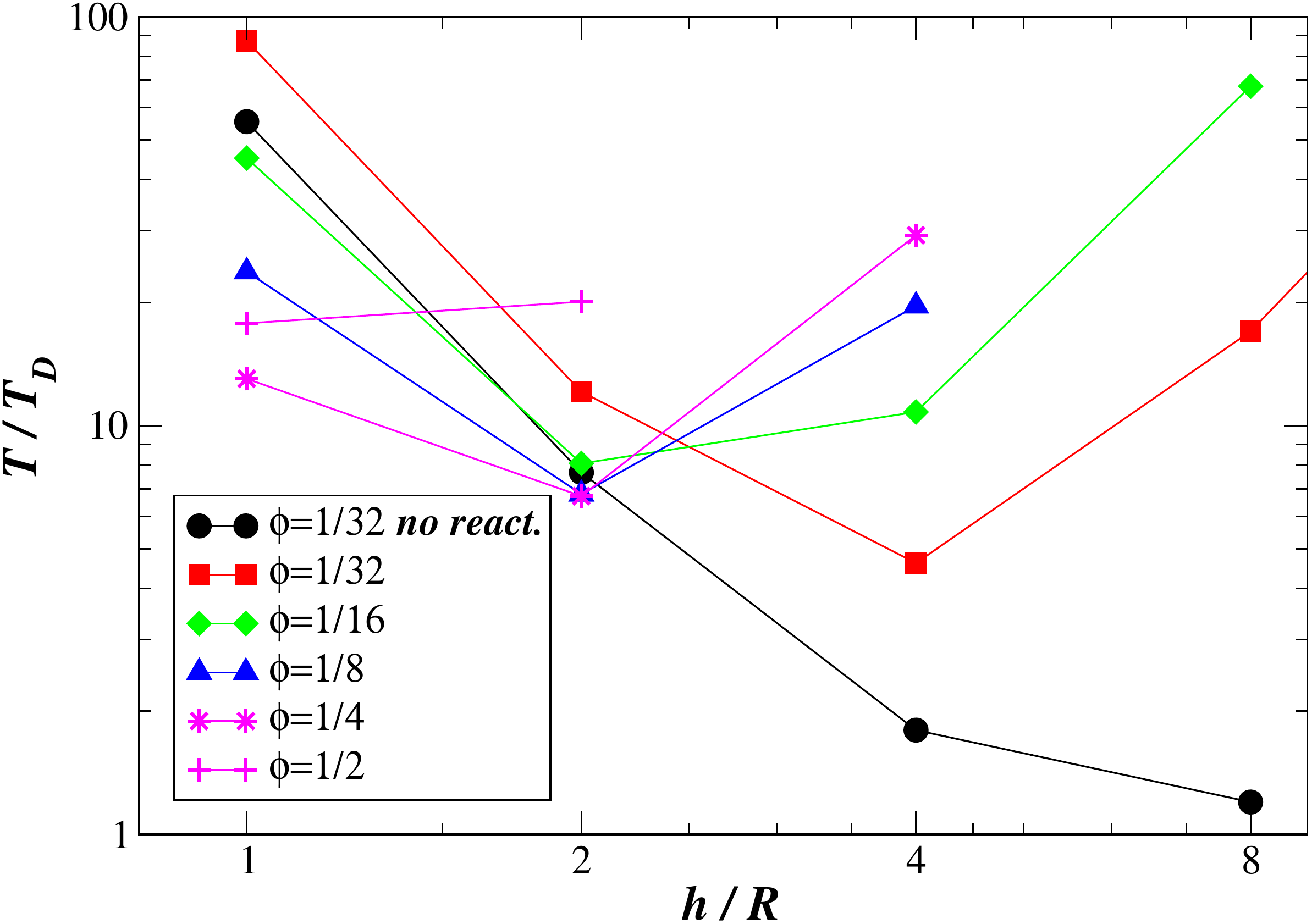}
\par\end{centering}
\caption{\label{fig:OptimalGrid}Ratio of SRBD computational time $T$ needed
to simulate a reaction-diffusion process and the computational time
$T_{D}$ needed to simulate diffusion-only process, as a function
of cell size relative to reactive radius, for annihilation $A+B\rightarrow B$
and $\emptyset\rightarrow A$. We use the same parameters as used
to produce Fig. \ref{fig:MicroToMacro}, and set $\protect\D t=1$,
and employ system sizes ranging from 15.5 to 62 thousand particles.
For low densities, an optimal cell size is observed ranging from $h=4R$
to $h=2R$. For densities $\phi\geq1/2$, the minimum cost occurs
for $h=R$ and the optimal grid size is the smallest possible one.
For comparison, we include for $\phi=1/32$ results for the case when
the reaction rates are set to zero so that they are still scheduled
but never actually happen.}
\end{figure}

In Fig. \ref{fig:OptimalGrid} we show some empirical results on the
cost of the SRBD algorithm as a function of the grid spacing. As a
comparison, we use the computational time $T_{D}$ needed to diffuse
the particles only, without processing any reactions. When we set
the reaction rates to zero so that no reactions actually happen, managing
the LLCs and event queue used to process reactions in SRBD (even though
empty), increases the cost and it is optimal to set $h=L$, where
$L$ is the system size. However, when reactions do happen, we see
that the computational time is very sensitive to the choice of grid
size $h$ and there is an optimal value $h_{\text{opt}}$ that minimizes
the cost. As expected, for dilute systems it is best to set the cell
size to be larger than the reactive radius, $h_{\text{opt}}>R_{\max}$,
so that the cost is dominated by diffusion as particles try to find
reactive partners, rather than being dominated by managing the grid-based
data structures used in the SRBD algorithm. For larger densities however,
the optimal choice is to make the cells as small as possible, $h_{\text{opt}}=R_{\max}$.
The exact optimal value of $h$ will depend not only on details of
the computational implementation and system size, but also on the
reaction rates and densities in a nontrivial way, and we recommend
that empirical testing is the best way to choose the optimal cell
size in practice.

\subsection{\label{subsec:TailsABC}Long-time tails for reversible association
$A+B\leftrightarrow C$}

In this section we continue investigating diffusion-limited reactions
at thermodynamic equilibrium in two dimensions. We focus here on the
reversible association reaction $A+B\underset{k}{\rightarrow}C$ and
$C\underset{\tilde{k}}{\rightarrow}A+B$, where the ratio of the forward
and backward rate is chosen to give an equilibrium steady state with
average number densities $\av{n_{A}}=\av{n_{B}}=\av{n_{C}}=n$. In
fact, SRBD reaches not just a steady state but a time-reversible state
of true thermodynamic equilibrium. Recall that the microscopic association
and dissociation mechanisms chosen in our Doi model formulation (see
Section \ref{subsec:DetailedBalance}) ensure detailed balance with
respect to a uniformly mixed equilibrium distribution. This means
that the steady state is a thermodynamic equilibrium state in which
the $A$, $B$ and $C$ molecules are uniformly mixed in the domain
and uncorrelated with one another. This implies that the forward reaction
rate for the association is $k=\pi R_{AB}^{2}\lambda$ as if the system
were reaction-limited and thus locally well-mixed. This holds \emph{independently}
of the value of $\lambda$, i.e., independently of whether the reaction
is actually diffusion-limited or reaction-limited.

However, the unusual macroscopic behavior of the system for diffusion-limited
parameters becomes evident if one considers not the static or instantaneous
rate, but rather, the dynamics of the fluctuations around the equilibrium
values. In particular, it has been known for some time that the microscopic
reaction mechanism affects the autocorrelation function (ACF) of the
fluctuations in the number density,
\[
\text{ACF}(t)=\av{\left(n_{C}(\tau+t)-n\right)\left(n_{C}(\tau)-n\right)},
\]
where $n_{C}(t)=N_{C}(t)/V$ is the instantaneous number density averaged
over the spatial extent of the domain. If the system were reaction-limited
and the usual law of mass action kinetics applied, the ACF would show
exponential decay, $\text{ACF}\left(t\right)\sim\exp\left(-3knt\right)$.
However, when the reaction is diffusion-limited, one observes a \emph{long-time
power-law tail} in the ACF \cite{LongTimeTails_ABC,AsymptoticPowerLaw_ABC},
\begin{equation}
\text{ACF}\left(t\right)\approx\frac{5}{216n\pi Dt}\,\text{ACF}\left(0\right),\label{eq:ACF_tail}
\end{equation}
where for simplicity we have assumed equal diffusion coefficient for
all species, $D_{A/B/C}=D$. The fundamental physics behind this long-time
tail is the fact that the reaction locally conserves $\tilde{n}=n_{A}+n_{B}+2n_{C}$,
which can only relax by slow diffusion. It is important to note that
the asymptotic behavior (\ref{eq:ACF_tail}) is \emph{independent}
of the reaction rate, i.e., independent of $\lambda$. Therefore,
the ACF switches at some characteristic time $\tau$ from exponential
behavior $\text{ACF}\left(t\ll\tau\right)\sim\exp\left(-3knt\right)$
with decay rate independent of $\lambda$, to an inverse time decay
(\ref{eq:ACF_tail}) with coefficient independent of $\lambda$ at
long times. The value of $\lambda$ determines the value of $\tau$.
For slow reactions, i.e., the reaction-limited case, exponential decay
dominates over most of the decay. For fast reactions, i.e., the diffusion-limited
case, the exponential decay lasts for only a very short amount of
time, and the majority of the decay of the ACF is much slower than
exponential.

\begin{figure}
\begin{centering}
\includegraphics[width=0.7\textwidth]{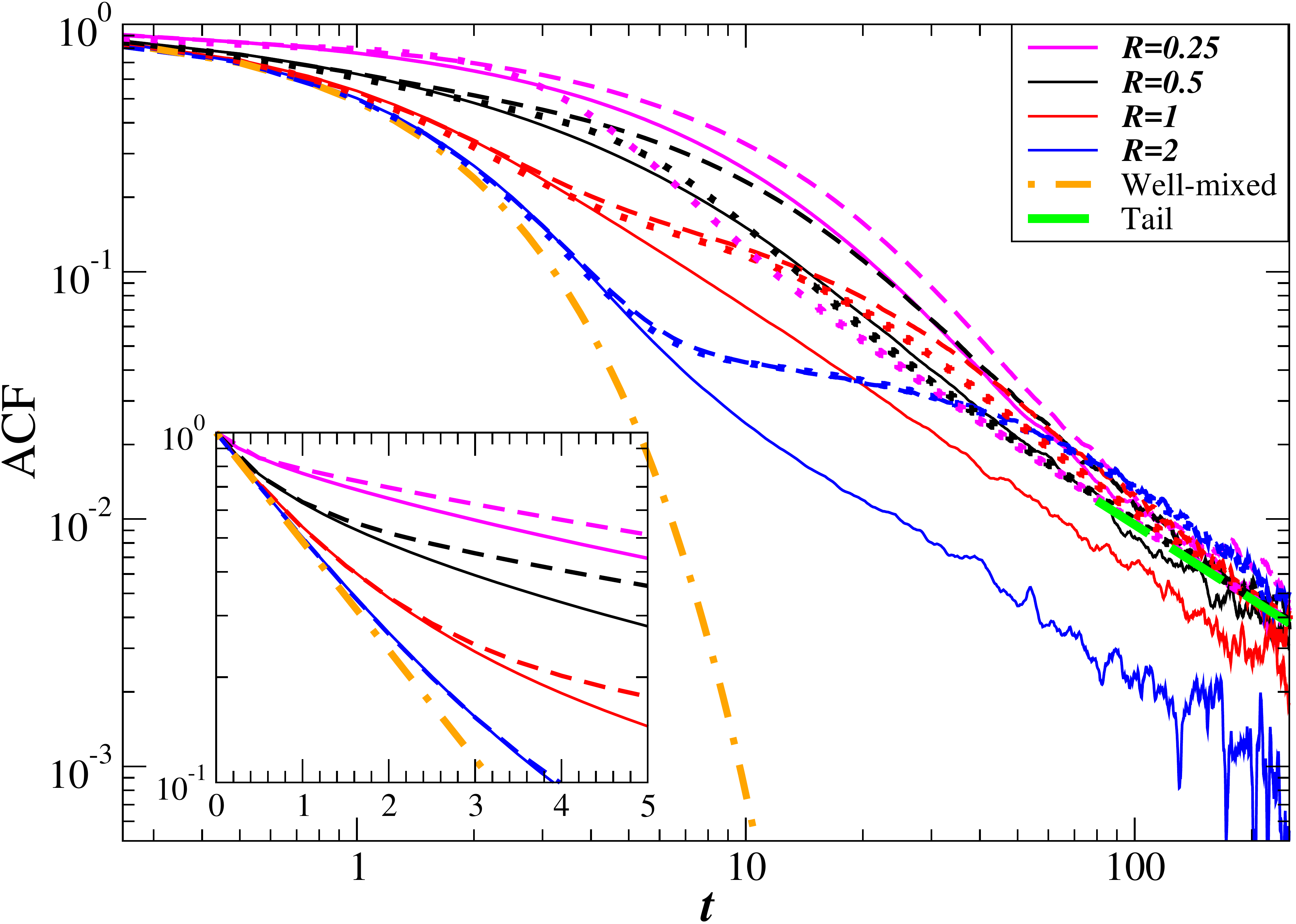}
\par\end{centering}
\caption{\label{fig:ABC-2D}Autocorrelation function of the total number of
$C$ molecules in a two-dimensional system undergoing the reversible
association $A+B\leftrightarrow C$. Solid lines show results for
SRBD with different values for the reactive radius $R$ (see legend),
and dashed lines of the same color are for RDME with reactive grid
spacing $h=2R$. The dotted lines are theoretical predictions for
RDME based on fluctuating hydrodynamics \cite{LongTimeTails_ABC}.
The thick dashed dotted orange line shows the exponential decay predicted
for a perfectly mixed system, and the thick dashed green line shows
the theoretical tail (\ref{eq:ACF_tail}). The inset focuses on early
times and uses a linear scaling of the $x$ axis in order to emphasize
the exponential decay observed in reaction-limited systems. Note that
the long-time tail has non-negligible statistical noise compared to
the signal when the ACF drops below $10^{-2}$.}
\end{figure}

In Fig. \ref{fig:ABC-2D} we show the ACF for SRBD (solid lines) for
several values of the reactive radius $R=R_{AB}=2R_{A}=2R_{B}$, for
$n=0.23873$ particles per unit (two-dimensional) area, $D=0.032549$,
$k=1$, $\tilde{k}=n$, for a square domain of length $L=64$ and
time step size $\D t=0.25$. For each $R$, we show in the same color
for comparison RDME (dashed lines) with a grid spacing $h=2R=4R_{A/B}$,
which was determined empirically to lead to an excellent matching
between SRBD and RDME at short to intermediate times. The inset in
the figure focuses on short times and shows that $\text{ACF}\left(t\rightarrow0^{+}\right)$
decays exponentially as predicted for a well-mixed (reaction-limited)
system (orange dashed-dotted line). This is because SRBD reaches a
steady state of thermodynamic equilibrium consistent with an ideal
solution (gas) of $A$, $B$ and $C$ molecules. We have confirmed
(not shown) that the two-particle (second order) correlation functions
of our system are consistent with an ideal gas mixture to within statistical
uncertainty.

For later times however, all of the ACFs shown in Fig. \ref{fig:ABC-2D}
show a much slower $t^{-1}$ decay than predicted for a well-mixed
system. For RDME we see that the ACF has the same universal tail given
by (\ref{eq:ACF_tail}), shown as a thick dashed green line, independent
of the value of the grid spacing $h$, as predicted by the theory.
However, for SRBD, different values of the reactive distance $R$
lead to the same $t^{-1}$ decay at long times, but with a different
coefficient. Empirically we find that fitting the SRBD tails with
(\ref{eq:ACF_tail}) gives an effective diffusion coefficient $D_{\text{eff}}=D+0.13knR^{2}.$
As already discussed in Section \ref{subsec:ExtraDiffusion}, this
enhancement of the effective diffusion coefficient by reactions comes
because a sequence of association/dissociation reactions in SRBD leads
to displacements of $A$ and $B$ molecules. Quite generally, we expect
to see enhancement of the diffusion coefficients on the order of $\sim R^{2}/\tau$,
where $\tau$ is a reactive time for the sequence $A+B\rightarrow C\rightarrow A+B$,
but the exact dependence is difficult to compute analytically.

In \cite{LongTimeTails_ABC,AsymptoticPowerLaw_ABC}, Gopich \emph{et
al.} developed a perturbative (one-loop) renormalization theory for
the tail of the ACF based on fluctuating hydrodynamics. In this theory,
one assumes that the fluctuations are weak and solves the fluctuating
hydrodynamics equations linearized around the equilibrium state, and
then evaluates the nonlinear term due to the binary reactions to quadratic
order in the fluctuations in order to estimate the leading-order corrections
due to fluctuations. The theory is a continuum theory, but can be
easily modified to account for the spatial discretization in RDME.
Specifically, we have replaced the spectral (negative) Laplacian $q^{2}$
with the modified (negative) Laplacian $\sin^{2}\left(qh/2\right)/\left(h/2\right)^{2}$
in Eq. (3) in \cite{LongTimeTails_ABC}, and replaced the integral
over wavenumbers $\V q$ in Eq. (11) in \cite{LongTimeTails_ABC}
with a sum over the discrete Fourier modes supported on the periodic
RDME grid. The integral over time in Eq. (11) in \cite{LongTimeTails_ABC}
can be performed analytically, and the resulting sum over wavenumbers
evaluated numerically. This gives a complete theory for $\text{ACF}(t)$,
not just the tail computed explicitly in \cite{LongTimeTails_ABC}.
In Fig. \ref{fig:ABC-2D} we show this theory with dotted lines for
different values of the RDME grid spacing $h$. The improved theory
is found to be in excellent agreement with the numerical data for
$h=4$ (dark blue lines), with the agreement becoming progressively
worse for smaller $h$, i.e., as the system becomes more and more
diffusion-limited. By increasing the number density, i.e., the number
of molecules per cell, we have confirmed that the mismatch with the
theory for intermediate times for diffusion-limited reactions cannot
be blamed on the fact that the fluctuations are not weak enough for
the perturbation analysis to apply. Instead, it appears that the theory
is missing some of the microscopic correlations that develop in diffusion-limited
systems, revealing once again the difficulty of quantitative continuum
modeling of such systems.

\subsection{\label{subsec:TuringPatterns}Pattern formation}

\begin{figure}
\begin{centering}
\includegraphics[width=0.75\textwidth]{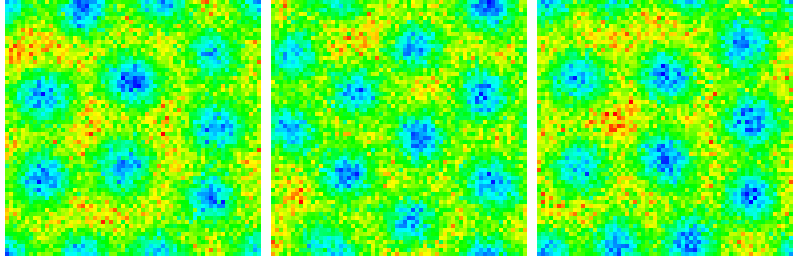}
\par\end{centering}
\caption{\label{fig:TuringBPM}Typical Turing-like quasi-steady patterns for
the BPM model at time $t=10^{4}$, illustrated here by color plots
of the number densities for $U$ molecules. Left panel is for RDME,
middle panel is for S-BD-RME, and right panel is for SRBD. The reactive
distance is set to $R=h=0.125$, i.e., a $256^{2}$ grid, but the
images were produced using a $64^{2}$ grid to compute local number
densities, in order to reduce the fluctuations.}
\end{figure}

In this section we study a two-dimensional reaction-limited system
with three species $U$, $V$ and $W$, undergoing seven reactions
according to the Baras-Pearson-Mansour (BPM) model \cite{Baras1990,Baras1996},
\begin{align*}
U+W & \rightarrow V+W\\
V+V & \leftrightarrows W\\
V & \leftrightarrows0\\
U & \leftrightarrows0.
\end{align*}
The reaction rates are chosen to give a limit cycle for the reaction
in the absence of diffusion, and the diffusion coefficient of the
$U$ species is chosen to be very different, $D_{V}=D_{W}=D_{U}/10$,
which leads to the formation of Turing-like spot patterns forming
a hexagonal or a monoclinic structure (see Fig. 5 in \cite{FluctReactDiff}).
A typical quasi steady-state pattern obtained for the parameters we
use is illustrated in Fig. \ref{fig:TuringBPM}.

This system was studied using the RDME and fluctuating hydrodynamics
by some of us in \cite{FluctReactDiff}, and it was concluded that
fluctuations accelerate the initial formation of a disordered spot
pattern, and also accelerate the subsequent annealing of defects to
form a lattice of spots. Here we repeat the same computations but
using SRBD and S-BD-RME, in order to understand the importance of
the microscopic model to the pattern formation. We have studied systems
that are either initialized in a uniformly mixed state corresponding
to a point on the limit cycle, which leads to initial oscillations
of the average concentrations which get damped until a fixed Turing
pattern is established, as well as systems initialized to be at the
unstable fixed point of the limit cycle, so that fluctuations kick
it onto the limit cycle via growing oscillations until eventually
a fixed pattern forms. Here we focus on the setup studied in \cite{FluctReactDiff}
and initialize the system on the limit cycle; initially the particles
of all three species are uniformly and randomly distributed throughout
the domain.

We use the same reaction parameters as reported in Section VB in \cite{FluctReactDiff}.
Although the system is two-dimensional we think of it as a three-dimensional
system with a small thickness $\D{z=0.5}$ in the third dimension
\footnote{Note that the thickness is denoted with $A$ in \cite{FluctReactDiff}.},
so that number densities are still expressed in units of particles
per unit volume rather than per unit area. For simplicity, in SRBD
we set $R_{U}=R_{V}=R_{W}=R/2$, where $R$ is the reactive distance.
The chosen reaction rates and diffusion coefficients are such that
the system is reaction-limited, so that we can easily obtain the effective
macroscopic reaction rate for any binary reaction from the two-dimensional
equivalent of (\ref{eq:lambda_to_k_mix_3D}) \footnote{In one dimension the corresponding formula would be $k=s\left(2\pi R\,\D y\,\D z\right)\lambda$.},
\begin{equation}
k=s\left(\pi R^{2}\D z\right)\lambda.\label{eq:lambda_to_k_mix_2D}
\end{equation}
This allows to obtain $\lambda$ for each reaction from the knowledge
of the desired effective reaction rate $k$ for each of the three
binary reactions in the BPM model. For S-BD-RME and RDME, we can simply
use the desired effective rates $k$ as input microscopic rates, independent
of the reactive cell size $h$. We vary $R$ and $h$ while keeping
the overall system size fixed at $L_{x}=L_{y}=32$.

In SRBD we use the minimal possible grid spacing $h=R$ for sampling
reactions, in order to maximize computational efficiency; in what
follows we will use $h$ when comparing SRBD and RDME/S-BD-RME simulations
but we remind the reader that $h$ has no physical meaning for SRBD.
We change the grid size $N_{x}=N_{y}=L_{x}/h$ from $N_{x}=64$ to
$N_{x}=512$; for the chosen rate parameters we would need grids larger
than $1024\times1024$ to enter the diffusion-limited regime. Our
RDME computations are performed using the split scheme described in
Appendix A of \cite{FluctReactDiff}. The time step size $\D t$ for
\emph{all} methods is limited by the fast diffusion of $U$ molecules,
and we set the diffusive Courant number to $D_{U}\D t/h^{2}\approx0.3$.
The total number of molecules (particles) in the system can be as
large as $2.5\cdot10^{6}\D z$, i.e., about a million particles for
our setup. As a rough idea of the computational effort involved, we
note that for $N_{x}=256$, the total running time up to physical
time $T=10^{4}$ on a 3GHz Intel Xeon processor was 4.5h for RDME,
12h for S-BD-RME, and 19h for SRBD. For $N_{x}=512$, the running
time was 54h for S-BD-RME and 48h for SRBD, which decreased to 43h
when a $256^{2}$ grid was used to process the reactions in SRBD (see
discussion in Section \ref{subsec:OptimalGrid}).

\begin{figure}
\begin{centering}
\includegraphics[width=0.49\textwidth]{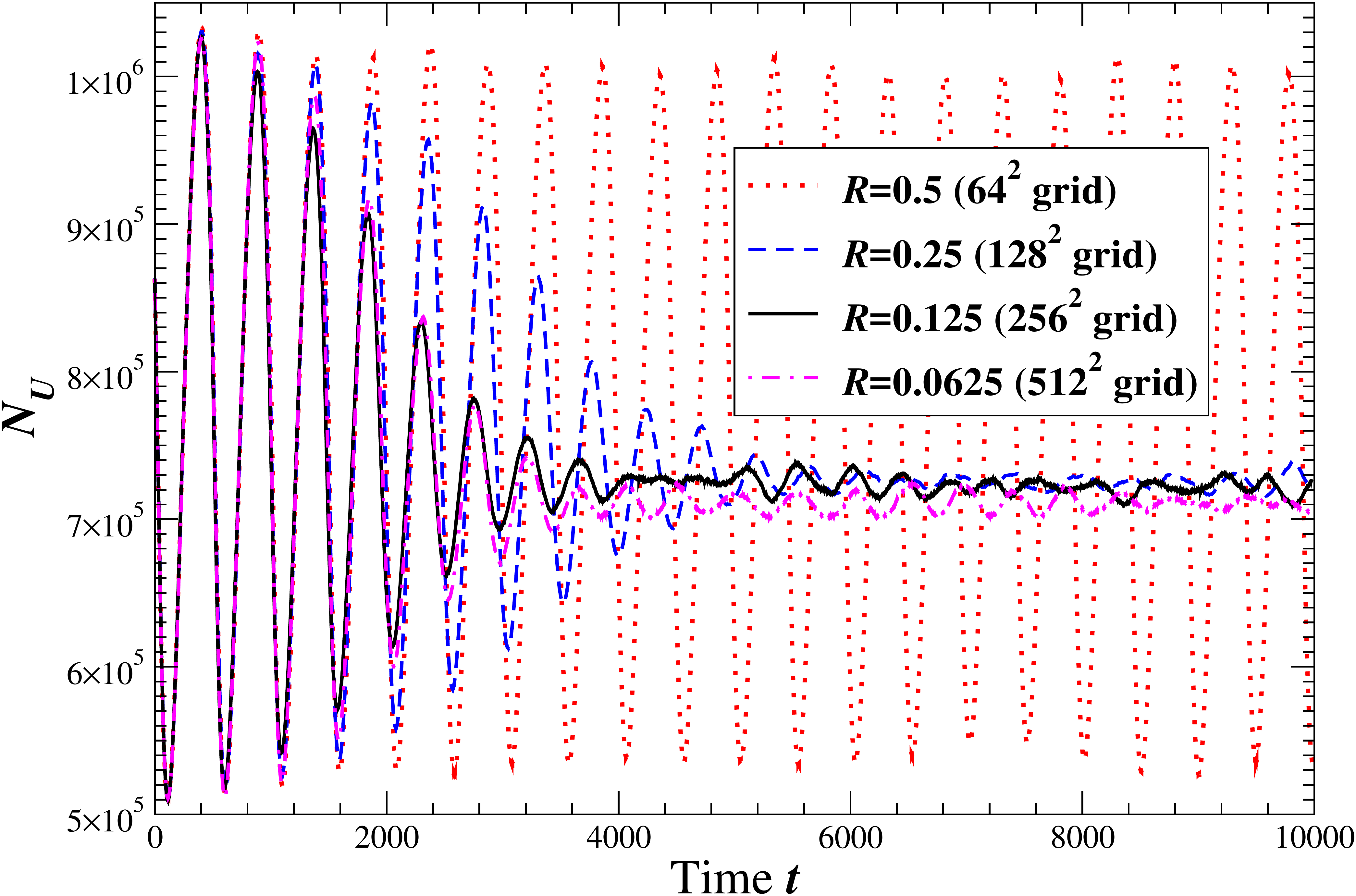}\includegraphics[width=0.49\textwidth]{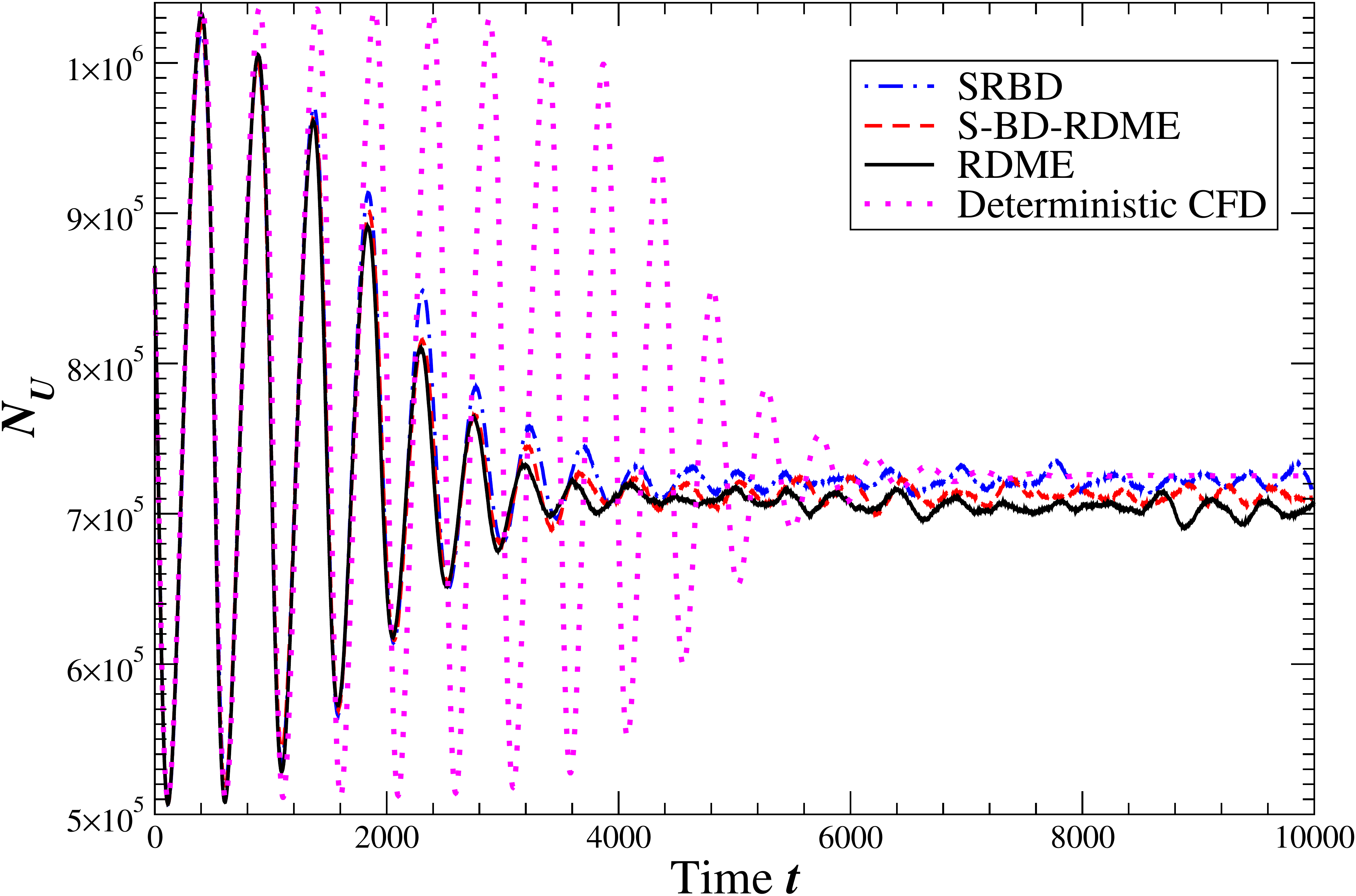}
\par\end{centering}
\caption{\label{fig:BPM_2D_resolution}Total number $N_{U}\left(t\right)$
of $U$ molecules in a pattern-forming BPM reaction-diffusion model
in two dimensions. Initially $N_{U}$ oscillates on the limit cycle
of the reaction-only model, until a Turing-like spatial instability
leads to the formation of a quasi-steady spot pattern. (\emph{Left})
Results for SRBD for different values of the reactive distance $R$
(see legend). The computations used a grid (size shown in legend)
of spacing $h=R$ for processing reactions. (\emph{Right}) Comparison
between SRBD with $R=0.125$ to S-BD-RME and RDME with reactive grid
spacing $h=R=0.125$. The pattern formation is delayed for deterministic
CFD \cite{FluctReactDiff} on a grid of spacing $h=0.125$, initialized
with a statistically indistinguishable initial condition.}
\end{figure}

In Fig. \ref{fig:BPM_2D_resolution}, we show the total number of
$U$ molecules as a function of physical time. This number oscillates
according to the limit cycle for a while, until a quasi-steady pattern
is formed. The types of ``steady'' Turing-like patterns obtained
using SRBD and S-BD-RME are visually indistinguishable from those
obtained using RDME, as illustrated in Fig. \ref{fig:TuringBPM}.
In the presence of fluctuations the final patterns are not strictly
stationary and the Turing spots can diffuse, however, this happens
on a slow time scale not studied here. Of main interest to us is the
typical time it takes for the Turing pattern to emerge.

In the left panel of Fig. \ref{fig:BPM_2D_resolution}, we show results
for SRBD for four different values of the particle diameter $R$.
Somewhat unexpectedly, the results for S-BD-RME with grid spacing
$h=R$ were found to be visually indistinguishable from those for
SRBD with reactive radius $R$. For SRBD or S-BD-RME, we do not see
a Turing pattern emerging even after $t=10^{5}$ when $R=0.5$. By
contrast, a pattern is formed in less time even if the deterministic
reaction-diffusion equations (corresponding to the limit $\D z\rightarrow\infty$)
are solved using standard computational fluid dynamics (CFD) techniques
\cite{FluctReactDiff} starting from a random initial condition. A
pattern does form for SRBD for $R=0.25$, and it forms faster yet
for $R=0.125$, and no noticeable change happens when we reduce the
reactive distance even further to $R=0.0625$. Somewhat surprisingly,
the results for RDME were found to be rather \emph{independent} of
the grid size, as also observed in Fig. 7 in \cite{FluctReactDiff},
and are statistically hard to distinguish from those obtained using
SRBD or S-BD-RME for $R\leq0.125$, as illustrated in the right panel
of Fig. \ref{fig:BPM_2D_resolution}.

We can explain the difference between RDME and SRBD/S-BD-RME by the
observation that diffusion is enhanced by reactions in methods in
which diffusion takes place off lattice, as we discussed in Section
\ref{subsec:ExtraDiffusion} and quantified in Section \ref{subsec:TailsABC}.
Indeed, the formation of the pattern can be suppressed also by enlarging
the diffusion coefficients of the particles in RDME. Since the effective
enlargement of the diffusion coefficients in SRBD/S-BD-RME is proportional
to $R^{2}$, using smaller reactive radii makes the effect smaller.

\begin{figure}
\begin{centering}
\includegraphics[width=0.7\textwidth]{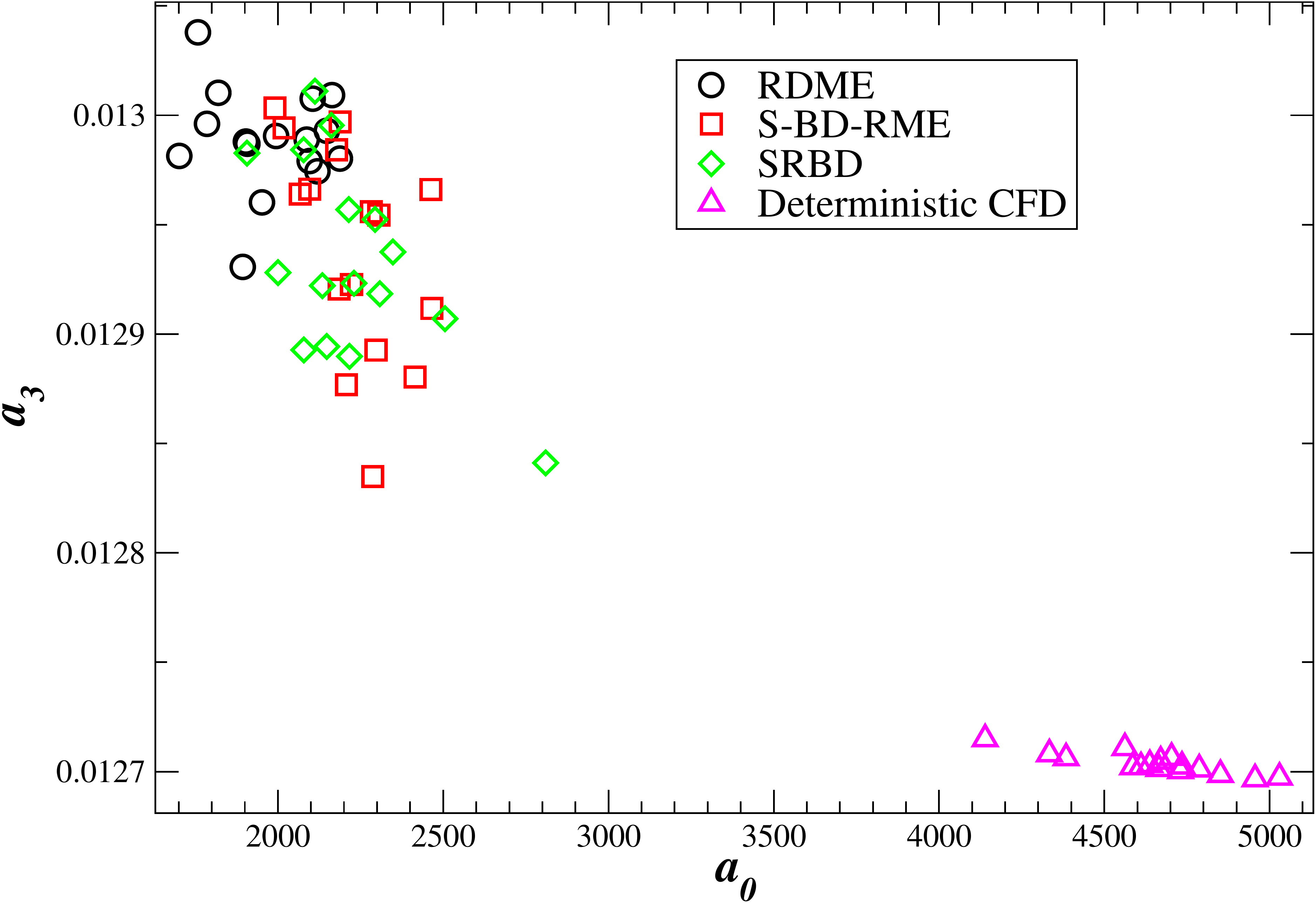}
\par\end{centering}
\caption{\label{fig:TuringFits}Scatter plot of the fitting coefficients $a_{0}$
and $a_{3}$ in the empirical fit (\ref{eq:N_U_fit}) for the dynamics
of pattern formation in the BPM model, for 16 statistically independent
simulations using one of four different methods (see legend).}
\end{figure}

We can make the comparison between the different methods for $R=0.125$
more quantitative by fitting the total number of $U$ molecules as
a function of time to the empirical fit (see Eq. (41) in \cite{FluctReactDiff})
\begin{equation}
N_{U}(t)=\left(1-\tanh\left((t-a_{0})/a_{2}\right)\right)\left(a_{1}\,\sin(a_{3}t+a_{4})+a_{5}\right)+a_{6},\label{eq:N_U_fit}
\end{equation}
and comparing the distribution of the fitting parameters for the different
methods over a set of statistically-independent runs. This comparison
is shown in Fig. \ref{fig:TuringFits}. In this figure we show the
values of the fitting parameters $a_{0}$, which represents the onset
time for the pattern formation, and $a_{3}$, which represents the
frequency of the oscillation, for 16 samples initialized using statistically
independent random configurations and using different random number
streams, for RDME, S-BD-RME, SRBD and deterministic CFD. Each of the
methods forms a cluster in this plane, and it is clear that deterministic
CFD is quite distinct in both the onset time and the frequency (note,
however, that the range of the $y$ axis is rather small) from the
methods that account for fluctuations. All methods that include fluctuations
are relatively similar in this comparison, but the RDME cluster is
seen to be somewhat separated from SRBD and S-BD-RME, which are themselves
not distinguishable in this statistical test. We believe that the
small difference between RDME and SRBD/S-BD-RME stems from the slightly
enhanced diffusion in the Doi model compared to lattice-based models.

\section{\label{sec:Conclusion}Conclusions and Future Directions}

We described a novel method for simulating the Doi or volume reactivity
model of reaction-diffusion systems. The SRBD algorithm is based on
time splitting of diffusion and reaction, and uses an event-driven
algorithm to schedule and process reactions during a time step without
any approximations. This makes the method robust from the reaction-limited
to the diffusion-limited case, and allows one to easily control the
numerical error in SRBD by reducing the time step size. Unlike simpler
algorithms for incorporating thermal fluctuations in reaction-diffusion
models, such as the widely-used RDME, a grid is only used in SRBD
to dramatically improve computational efficiency, without affecting
physical observables. The SRBD method is therefore a true particle
method that maintains Galilean invariance and isotropy, just like
molecular dynamics.

Our studies of irreversible association in Section \ref{subsec:MicroToMacro}
showed the complexity of the conversion from microscopic to effective
macroscopic rates for diffusion-limited reactions. Such conversion
is in fact not possible in two dimensions, and even in three dimensions
the corrections to the effective rate depend non-analytically on the
density \cite{Coagulation_Renormalization_PRL,Coagulation_Renormalization}.
This brings into question the law of mass action kinetics for diffusion-limited
systems, even at macroscopic length and time scales, casting doubts
on any attempt to model such systems using local in space and time
reaction-diffusion equations. Our studies of diffusion-limited reversible
association in Section \ref{subsec:TailsABC} indicated that even
though a simple law of mass action kinetics can model the \emph{instantaneous}
macroscopic reaction rate, long-lived temporal correlations of the
fluctuations lead to effective memory in the dynamics \cite{RelaxationTime_ABC}.
Furthermore, we numerically demonstrated that by ensuring reversibility
of the microscopic reaction rules we achieve a state of true thermodynamic
equilibrium \cite{DiffusionLimited_NonDB}.

For reaction-limited systems, diffusion is fast enough to mix the
reactants at microscopic and mesoscopic scales, and the conversion
from microscopic to macroscopic rates is much simpler. Indeed, in
Section \ref{subsec:TuringPatterns} we found a good matching between
the simpler and more efficient RDME model and SRBD/S-BD-RDME for sufficiently
small reactive distances, for a system undergoing a pattern-forming
Turing-like instability. Nevertheless, we found that when finite-range
reactions are combined with off-lattice diffusion, reversible reactions
increase the effective diffusion coefficient by an amount on the order
of $R^{2}/\tau$, where $R$ is the typical reactive distance, and
$\tau$ is the typical duration of a binding-unbinding sequence. 

One can argue that the enhancement of the diffusion coefficients by
reactions that we observed in the Doi model is unphysical, since one
considers reaction and diffusion to be separate physical processes
that cannot couple by the Curie principle. At the same time, for diffusion-limited
systems reaction and diffusion are intimately coupled, and it is not
in fact obvious that the traditional reaction-diffusion partial differential
equation, in which reaction and diffusion are completely decoupled,
is appropriate (even once rates are renormalized). Since diffusion
affects macroscopic reaction rates, it is perhaps natural to expect
that reactions should in turn affect diffusion on macroscopic scales.

The enhanced diffusion in the Doi model occurs because of the simplifications
of actual reactive mechanisms. Notably, all molecules are considered
to be point-like particles and only identified by a species label.
When a forward reaction $A+B\rightarrow C$ takes place, all information
about the original reactant particles is lost, and is subsequently
stochastically re-created by the reverse reaction. This is equivalent
to saying that once the $C\equiv AB$ complex forms, it rotates and
rearranges internally on a much faster time scale than it diffuses,
which is unphysical. In an actual reaction, however, the $A$ or $B$
will denote atomic units or molecular subunits that will retain their
identity via the reaction, and a more physically-realistic model may
treat the $C$ as a combination $A+B$ of units that are bonded by
an elastic or rigid bond, thus retaining the rotational diffusion
of the $C$ molecule (molecular complex), and the finite time scales
of the internal dynamics of this complex. Developing more physically
realistic microscopic models is well beyond the scope of the present
work, but remains an important topic for future study. It should be
noted that the key algorithmic ideas developed in this work can find
use well beyond just the specific Doi model we employed in this work.

A key disadvantage of event-driven algorithms such as FPKMC and SRBD
is the difficulty of parallelization of event loops without making
uncontrolled approximations. The diffusion step in SRBD can be trivially
parallelized since each particle diffuses independently of other particles.
The parallelizaton of the reactive step, however, requires sophisticated
``time-warp'' technology only recently employed for kinetic Monte
Carlo simulation \cite{Parallel_KMC_Bulatov}. Future work should
explore whether it is possible to parallelize SRBD using simpler techniques
by using the fact that the time step size provides an upper bound
on the maximum time to the next event.

In the SRBD model used in this work, we assumed that the particles
are independent Brownian walkers. It is well-known that this model
is not appropriate for particles diffusing in a liquid, because of
the importance of long-ranged hydrodynamic correlations (often called
hydrodynamic interactions) mediated by momentum transport in the solvent.
The time splitting used in SRBD makes it very easy to use recently-developed
linear-scaling Brownian Dynamics with Hydrodynamic Interactions (BD-HI)
\cite{SpectralRPY} to diffuse particles. This would enable studies
of the importance of hydrodynamics to reaction-diffusion processes
in crowded fluid environments such as the cell cytoplasm.

While the SRBD method is most powerful and efficient at higher densities,
by changing the size of the grid used to accelerate the reaction handling
one can also reasonably efficiently handle lower densities, as shown
in Section \ref{subsec:OptimalGrid}. Nevertheless, at very low densities
the majority of the computing effort will be spent diffusing molecules,
which will take a long time to find other molecules to react with.
The problem of diffusing molecules rapidly over large distances without
incurring any approximation errors is already elegantly solved by
the FPKMC algorithm \cite{FPKMC_PRL,FPKMC1,FPKMC2}. It is in fact
possible to combine the FPKMC idea of propagating particles inside
protective regions when they are far from other particles they could
react with, with the SRBD handling for particles when they come close
to each other. This can be accomplished relatively easily by adding
to FPKMC a new kind of event, an SRBD time step, which is scheduled
in regular intervals of $\D t$ to update those particles not protected
by a domain by a time step of the SRBD algorithm. This kind of algorithm
obviates the need for the complex handling of particle pairs in FPKMC
while still retaining the key speedup of the FPKMC algorithm, and
works naturally with the Doi reaction model even in regions of high
densities. We leave such generalizations of the SRBD algorithm for
future work.
\begin{acknowledgments}
We thank Samuel Isaacson, Radek Erban, Martin Robinson, Charles Peskin,
and Alejandro Garcia for numerous stimulating and informative discussions,
and an anonymous referee for suggesting alternative perspectives on
the observation that diffusion is enhanced by reactions in the Doi
model. A. Donev was supported in part by the U.S. Department of Energy
Office of Science, Office of Advanced Scientific Computing Research,
Applied Mathematics Program under Award Number DE-SC0008271. C.-Y.
Yang was supported by the Reed College President\textquoteright s
summer fellowship during his visit to the Courant Institute. C. Kim
was supported by the U.S. Department of Energy, Office of Science,
Office of Advanced Scientific Computing Research, Applied Mathematics
Program under Contract No. DE-AC02-05CH11231.
\end{acknowledgments}

\section*{Appendix}

\appendix

\section{\label{app:ReactionRules}Microscopic Reaction Rules}

In this work, we adopt the following microscopic reaction rules based
on classifying each reaction into one of several categories:
\begin{enumerate}
\item \textbf{Death}: $A\rightarrow\emptyset$. Every particle of species
$A$ can disappear with rate $\lambda$ per unit time.
\item \textbf{Birth}: $\emptyset\rightarrow A$. With rate $\lambda$, a
particle $A$ is created randomly and uniformly inside the domain.
This reaction is the reverse of death.
\item \textbf{Conversion}: $A\rightarrow B$. Every particle of species
$A$ has a rate $\lambda$ of changing species into $B$. This reaction
is its own reverse (swapping the roles of $A$ and $B$).
\item \textbf{Annihilation}: $A+B\rightarrow\emptyset$ with rate $\lambda$
if within distance $R_{AB}$, where $B$ can be equal to $A$. Both
particles disappear.
\item \textbf{Production}: $\emptyset\rightarrow A+B$. The particle of
species $A$ is born randomly and uniformly in the domain (as for
birth), and the particle of species $B$ is born with a random position
uniformly distributed within a reactive sphere of radius $R_{AB}$
around the position of the particle of species $A$. This reaction
is the reverse of annihilation.
\item \textbf{Catalytic death}:\textbf{ }$A+B\rightarrow A$ with rate $\lambda$
if the $A$ and $B$ molecules are within distance $R_{AB}$. The
particle of species $B$ disappears.
\item \textbf{Catalytic birth}: $A\rightarrow A+B$. Every particle of species
$A$ has a rate $\lambda$ of splitting per unit time. A new particle
of species $B$ is born with a random position uniformly distributed
within a reactive sphere of radius $R_{AB}$ around the position of
the particle of species $A$. This reaction is the reverse of catalytic
death.
\item \textbf{Binding}: $A+B\rightarrow C$ with rate $\lambda$ if the
$A$ and $B$ molecules are within distance $R_{AB}$, where we allow
for $B$ to be equal to $A$, or for both $B$ and $C$ to be equal
to $A$ (for example, as in coagulation $2A\rightarrow A_{2}$). One
of the two reactants of species $A$ or $B$, chosen at random with
probability $1/2$, changes species into $C$, and the other reactant
particle disappears. Note that in specific applications it may be
more appropriate to always convert the $A$ into $C$ instead of the
random choice we implement here.
\item \textbf{Unbinding}: $A\rightarrow B+C$, where $B$ and/or $C$ can
be equal to $A$. Every particle of species $A$ has a rate $\lambda$
of splitting per unit time. The particle of species $A$ becomes a
product particle of species $B$ or $C$, chosen randomly with probability
$1/2$ (see comment for binding about changes of this rule), and the
second product particle is born randomly uniformly in a sphere centered
at the $A$ with radius $R_{BC}$. This reaction is the reverse of
coagulation.
\item \textbf{Catalytic conversion}: $A+B\rightarrow A+C$ with rate $\lambda$
if the $A$ and $B$ molecules are within distance $R_{AB}$, where
$C$ can be equal to $A$. The particle of species $B$ changes species
into $C$. This reaction is its own reverse.
\item \textbf{Transformation}: $A+B\rightarrow C+D$, with rate $\lambda$
if the $A$ and $B$ molecules are within distance $R_{AB}$. Here
$B$ can be equal $A$, and $C$ and $D$ can \emph{both} be equal
to $A$ (i.e., the case of catalysis is excluded). One of the two
reactant particles, chosen randomly with probability $1/2$, changes
species into $C$, and the other reactant particle changes species
into $D$. This reaction is its own reverse. Note that in specific
applications it may be more appropriate to always convert the $A$
into $C$ and the $B$ into a $D$ instead of the random choice we
implement here.
\end{enumerate}
We note that we have combined here rules that could be simplified
when some of the product/reactant species are identical for the sake
of brevity. For example, one could give a more condensed reaction
rule (and our code implements such condensed rules for efficiency)
for reactions such as $A+A\leftrightarrow A$ or $A+A\leftrightarrow A+B$.


\end{document}